\documentclass[aps,amsmath,twocolumn,amssymb,floatfixng,showpacs,
superscriptaddress,footinbib, longbibliography]{revtex4-1}

\usepackage{graphicx}% Include figure files
\usepackage{dcolumn}% Align table columns on decimal point
\usepackage{multirow}
\usepackage{booktabs}
\usepackage{bm,color}
\usepackage{braket}
\usepackage{amsmath,amssymb}
\usepackage[colorlinks,linkcolor=blue,hyperindex,CJKbookmarks]{hyperref}
\usepackage{epstopdf}
\usepackage{soul}
\usepackage{xcolor}

\newcommand{\bgea}{\begin{equation}}
\newcommand{\enea}{\end{equation}}
\newcommand{\bea}{\begin{eqnarray}}
\newcommand{\eea}{\end{eqnarray}}
\newcommand{\braOket}[3]{\langle n(#1)|#2|n(#3)\rangle}
\newcommand{\hatmu}{\hat\mu}
\newcommand{\U}{\mathop{\rm {}U}}

\begin{document}

\title{Topological Hall plateau in quasi-2D kagome magnet YMn$_6$Sn$_6$}

\author{Sambit Jena}
\affiliation{Department of Physics, École Centrale School of Engineering, Mahindra University, Hyderabad-500043, India}
\author{Nastaran Salehi}
\affiliation{Department of Physics and Astronomy, Uppsala University, Box 516, SE-75120 Uppsala, Sweden}
\author{Manuel Pereiro Lopez}
\affiliation{Department of Physics and Astronomy, Uppsala University, Box 516, SE-75120 Uppsala, Sweden}
\author{Olle Eriksson}
\affiliation{Department of Physics and Astronomy, Uppsala University, Box 516, SE-75120 Uppsala, Sweden}
\author{Narayan Mohanta}
\affiliation{Department of Physics, Indian Institute of Technology Roorkee, Roorkee 247667, India}
\author{Karthik V Raman}
\affiliation{Tata Institute of Fundamental Research, Hyderabad, Telangana 500046,  India}
\author{Tanay Nag}
\affiliation{Department of Physics, BITS Pilani-Hyderabad Campus, Telangana 500078, India}
\author{Banasree Sadhukhan}
\email{banasree.sadhukhan@mahindrauniversity.edu.in}
\affiliation{Department of Physics, École Centrale School of Engineering, Mahindra University, Hyderabad-500043, India}

%\date{\today}

\begin{abstract}

We examine the impact of the Dzyaloshinskii–Moriya interaction (DMI) in kagome magnets and show that a predominantly planar DMI together with ferromagnetic exchange stabilizes a disordered skyrmion phase in quasi-two-dimensional (2D) YMn$_6$Sn$_6$.  Within an {\it{ab initio}} framework combining density functional theory and spin-dynamics simulations, we generate realistic spin textures of disordered skyrmion and find that this phase persists for $B_{\rm ext} \leq 0.5$~T, with a decreasing skyrmion size as   magnetic field increases. We demonstrate the emergence of  topological Hall plateau  in the range $-0.5 \leq B_{\mathrm{ext}} \leq 0.5$~T, driven by nearly uniform scalar spin chirality and the resulting constant real-space Berry curvature. This response is anti-symmetric with magnetic field while magnitude and sign of these plateau are determined by a complex interplay between Hund’s coupling strength and chemical potential signifying the role of  Dirac points and van Hove singularities. In addition, we reveal topological magnon excitations in the disordered skyrmion phase of quasi-2D YMn$_6$Sn$_6$.

\end{abstract}

%%%%%%%%%%%%%%%%%%%%%%%%%%%%%%%%%%%%%%%%%%%%%%%%%%%%%%%%%%%%%%%%%%%%%%%%%%%%%%%%

\maketitle

%%%%%%%%%%%%%%%%%%%%%%%%%%%%%%%%%%%%%%%%%%%%%%%%%%%%%%%%%%%%%%%%%%%%%%%%%%%%%%%%%%%%

\section{Introduction} 
In recent times, nontrivial spin textures have become a central focus in condensed matter physics owing to their distinct topological characteristics and potential applications for next-generation technologies \cite{Nagaosa2013-xt, Rosler2006-xa, doi:10.1126/science.1166767,  Yu2010-ys,  Heinze2011-wt, Khanh2020-of, SciPostPhys.18.2.064, sadhukhan2025engineering}. Chiral domain walls, merons, bimerons, and skyrmions exemplify noncollinear magnetic configurations exhibiting enhanced stability against external perturbations due to their symmetry protection \cite{Nagaosa2013-xt,  fert2017magnetic,  ezawa2011compact}. Interestingly, magnetic skyrmions have attracted an intense interest because of their nanoscale dimensions, structural robustness, and efficient controllability via ultralow current densities. These features allow skyrmions as promising building blocks for energy-efficient spintronic applications and motivate continued efforts toward their controlled realization and manipulation in realistic material systems {\cite{Zhang_2023, mukherjee2026skyrmion, yang2024fundamentals, MUKHERJEE2025173036}.

\par The Dzyaloshinskii–Moriya interaction (DMI) constitutes an antisymmetric exchange mechanism that is central to the emergence and stabilization of chiral magnetic textures, including skyrmions and merons.  Originally, Dzyaloshinskii introduced this concept to explain weak ferromagnetism in antiferromagnets.  Later, Moriya provided a microscopic foundation based on Anderson’s superexchange theory \cite{Dzyaloshinskii1957,  Dzyaloshinskii1958, Nagaosa2013}. It is represented by a vector quantity that encodes both the strength, determined by the magnitude, and orientation of the interaction, with its direction constrained by crystal symmetry as captured by Moriya’s rules \cite{Moriya1960PRL,  Moriya1960PR}.  Note that the  magnitude of DMI is accessible through both phenomenological models and first-principles approaches \cite{PhysRevB.103.094410, PhysRevB.105.104418}. DMI microscopically originates from spin-orbit coupling (SOC) in systems with broken inversion symmetry, either in the bulk or at interfaces. It has been experimentally realized that in non-centrosymmetric B20 bulk compounds, including MnSi, FeGe, MnGe and FeCoSi give rise to intrinsic bulk DMI, breaking inversion symmetry, that stabilizes  skyrmion lattices \cite{PhysRevLett.106.156603, PhysRevLett.115.036602,  PhysRevLett.116.247201,  PhysRevLett.125.117204}.   Subsequently, skyrmions have also been demonstrated in interfacial multilayer heterostructures such as Ir(111)/Fe, Ta/CoFeB, and Pt/Co, where strong SOC at interfaces induces interfacial DMI, enabling the formation of nanoscale skyrmions \cite{heinze2011spontaneous,  moreau2016additive,  boulle2016room,  woo2016observation}.  In several compounds and heterostructures,  the DMI is intrinsically two-dimensional (2D) in nature \cite{doi:10.1126/sciadv.1600304,  Banerjee2013-lm,  PhysRevB.99.104402}.  More recently, Néel-type skyrmions stabilized by isotropic DMI have also been reported in 2D magnetic systems, further expanding the materials platform for exploring chiral spin textures in low-dimensional  systems \cite{PhysRevB.109.024420}. Therefore, DMI is indispensable for the predictive design and manipulation of chiral magnetism in quantum materials.

\par Now coming to the transport associated with chiral magnetic textures, the topological Hall effect (THE) arises from real-space Berry curvature associated with structures like skyrmions, providing a distinct mechanism  from conventional and anomalous Hall effects \cite{PhysRevB.107.L081110, saha2026quantized}. In noncollinear spin configurations, a finite scalar spin chirality \cite{PhysRevB.92.115417,  PhysRevB.80.054416,  doi:10.1126/sciadv.aap9962}  generates an effective electromagnetic field for itinerant electrons via the spin Berry phase.  As electrons traverse a skyrmion lattice, their spins adiabatically follow the local magnetization, accumulating a Berry phase proportional to the topological charge that produces an emergent magnetic field namely, Berry curvature. This effective Berry field deflects charge carriers transversely due to the presence of anomalous velocity, giving rise to the topological Hall response \cite{doi:10.1126/sciadv.aap9962,  doi:10.1126/sciadv.abq2765} where the group velocity of electrons is ignored. Note that various kinds of Hall responses have been an interesting topic of study \cite{PhysRevB.103.144308,PhysRevB.106.045424,PhysRevB.104.245122,PhysRevB.105.214307,PhysRevB.103.235154,Nag_2025,saha2026quantized, rkr4-p5n4,PhysRevB.104.115420,PhysRevB.107.245141,PhysRevB.106.045417}.

\par The phenomenon was first observed in MnSi \cite{PhysRevLett.102.186602} and has been extensively studied within phenomenological and microscopic frameworks \cite{PhysRevB.99.174425,  PhysRevLett.117.027202,  PhysRevB.98.195439,  PhysRevB.92.115417,  PhysRevB.104.174432,  PhysRevB.100.064429,  PhysRevB.102.064430,  PhysRevB.111.134433,  PhysRevX.15.011054,  PhysRevLett.102.186602}.  However, a quantitative, material-specific framework for theoretically understanding and predicting experimental observations is still lacking. This is because first-principles approaches that combine density functional theory (DFT) with spin dynamics are severely limited by the large system sizes \cite{PhysRevX.15.011054}, leaving it as a challenge for over a decade.  We address the question of whether the THE can be investigated in real kagome materials using an {\it{ab initio}} framework \cite{wills2010full,  SECCHI201561,  PhysRevB.61.8906,  PhysRevB.68.104436,  PhysRevB.79.045209} (see Appendix). More precisely, how do the real-space magnetic spin textures, obtained using magnetic exchange interactions from Liechtenstein formalism-based DFT technique and spin dynamics (SD) simulations, vary with the applied magnetic field? Can there be a quantized THE response for a certain type of spin texture?

\par In this letter, considering the quasi-2D YMn$_6$Sn$_6$ Kagome magnet, we find that the calculated DMI vectors are planar in nature, along with a ferromagnetic contribution coming from the isotropic part of the magnetic exchanges.  We identify a planar DMI-driven disordered skyrmion phase \cite{doi:10.1126/sciadv.aar7043} that persists up to external fields of $B_{\rm ext} \simeq 0.5$~T with decreasing skyrmion radius.  We show plateau-like Hall response, computed using the Kubo formalism,  within $-0.5 \leq B_\mathrm{ext} \leq 0.5$~T,  arising from real-space Berry curvature associated with nearly constant scalar spin chirality in this field window. The Dirac point and van Hove singularity can non-trivially change the conductivity plateau. Additionally,  we report topological magnons in the disordered skyrmion phase of quasi-2D YMn$_6$Sn$_6$.

\section{Effect of DMI in a Kagome Magnet}
 Bulk DMI,  arising from strong SOC in systems, plays a central role in stabilizing complex magnetic textures in kagome magnets. It is described by the antisymmetric exchange Hamiltonian, as given by
\begin{equation}
\mathcal{H}_{\text{bulk}}^{\rm DMI} = \sum_{\langle i,j \rangle} \mathbf{D}_{ij} \cdot \left( \mathbf{S}_i \times \mathbf{S}_j \right),
\end{equation}
where $\langle i,j \rangle$ denotes nearest-neighbor (NN) pairs, $\mathbf{S}_i$, $\mathbf{S}_j$ are spin vectors at sites $i$ and $j$ and $\mathbf{D}_{ij}$ is dictated by the underlying lattice symmetry.  For bulk DMI, all components of DMI vector $\mathbf{D}=(D_{x}$,  $D_{y}$, $D_{z}$ $\neq 0)$  can contribute.  In kagome systems, the bulk DMI typically possesses a relatively strong out-of-plane component $D_{z}$. This lifts the degeneracy of frustrated spin configurations and induces weak canting of spins away from collinear or coplanar arrangements. 

%%%%%%%%%%%%%%%%%%%%%%%%%%%%%%%%%%%%%%%%%%%%%%%%%%%%%%%%%%%%%%%%%%%%%%%%%%%%%%%
\begin{figure}[ht]
\centering
\includegraphics[width=0.5\textwidth,angle=0]{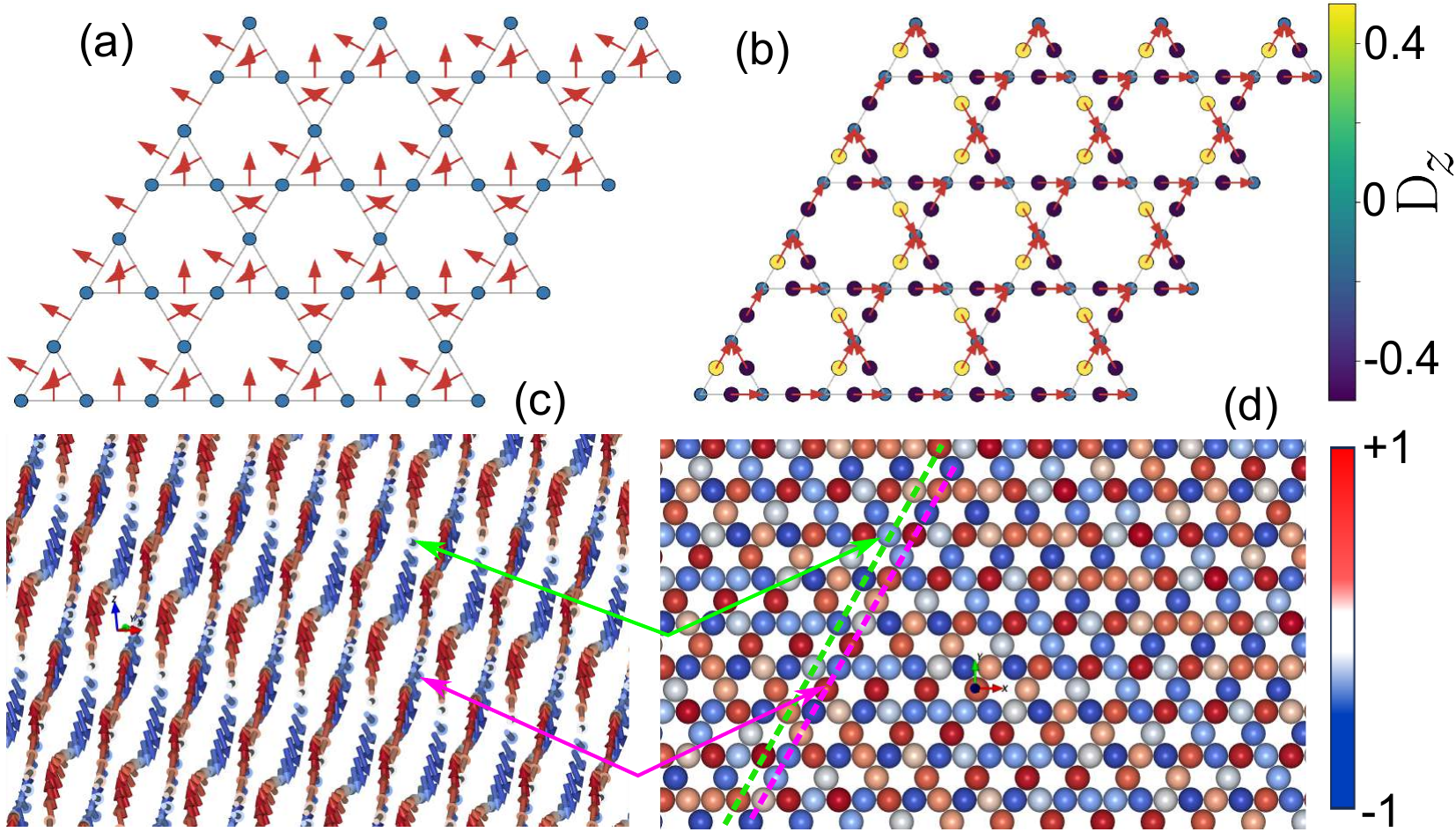}
\caption{(a) Planar and (b) bulk Dzyaloshinskii–Moriya interactions (DMIs) in kagome magnets.  (c)-(d) Combination of two intertwined spin spirals and a single spin spiral as ground state in a kagome magnet in presence of planar DMI.  The green and magenta colours represent the columns for intertwined spin spirals and a single spin spiral, respectively, in the kagome lattice mapping.}
\label{fig1}
\end{figure}

%%%%%%%%%%%%%%%%%%%%%%%%%%%%%%%%%%%%%%%%%%%%%%%%%%%%%%%%%%%%%%%%%%%%%%%%%%%%%%%

\par In contrast, planar DMI, originating from broken inversion symmetry at surfaces or in heterostructures,  is predominantly in-plane and energetically favors chiral twisting between neighboring spins separated by a distance $\hat{\mathbf{r}}_{ij}$ as given by $\mathbf{D}_{ij} = D_{ij} \left( \hat{\mathbf{z}} \times \hat{\mathbf{r}}_{ij} \right)$. Substituting this vector form into the DMI Hamiltonian,  we obtain 
\begin{equation}
\mathcal{H}_{\text{planar}}^{\rm DMI} = D_{ij} \sum_{\langle i,j \rangle}  \left( \hat{\mathbf{z}} \times \hat{\mathbf{r}}_{ij} \right) \cdot \left( \mathbf{S}_i \times \mathbf{S}_j \right). 
\end{equation}
Note that the planar DMI vector supports $\mathbf{D}=(D_{x}\ne 0$,  $D_{y} \ne 0$, $D_{z}= 0)$ with  strength  $|\mathbf{D}_{ij}| = \sqrt{D_x^2 + D_y^2}$.   This form of planar DMI favors chiral spin configurations, such as cycloidal spirals and skyrmions, depending on the lattice geometry and magnetic anisotropy.  This results in a finite scalar spin chirality
$\chi_{ijk} = \mathbf{S}_i \cdot \left( \mathbf{S}_j \times \mathbf{S}_k \right)$
and gives rise to emergent gauge fields affecting electronic transport.

\par To explore the above phenomena, we took a spin Hamiltonian for a kagome plane containing ferromagnetic exchange J$_{ij}$ = 1,  and D$_{x}$ = D$_{y}$ = 0.5.  The D$_{z}$ is zero for planar DMI whereas it is 0.4 for bulk DMI as shown in Fig.\ref{fig1} (a)-(b).  The spin texture has been simulated using SD simulation for both bulk and planar DMI cases.  In ferromagnetic kagome lattices,  planar DMI leads to the combination of two intertwined spin spirals and a single spin spiral with a well-defined handedness as shown in Fig.\ref{fig1} (c)-(d) whereas the bulk DMI leads to local spin canting only.   Compared to bulk DMI,  planar DMI is significantly more effective for kagome lattices in generating robust chiral spin textures, as it directly promotes long-wavelength twisting rather than merely inducing local canting.  As a result,  planar DMI provides a stronger driving force for stabilizing nontrivial topological textures such as spirals and skyrmions in kagome systems, offering a more efficient route to engineer chiral magnetism.

\begin{figure}[ht]
\centering
\includegraphics[width=0.5\textwidth,angle=0]{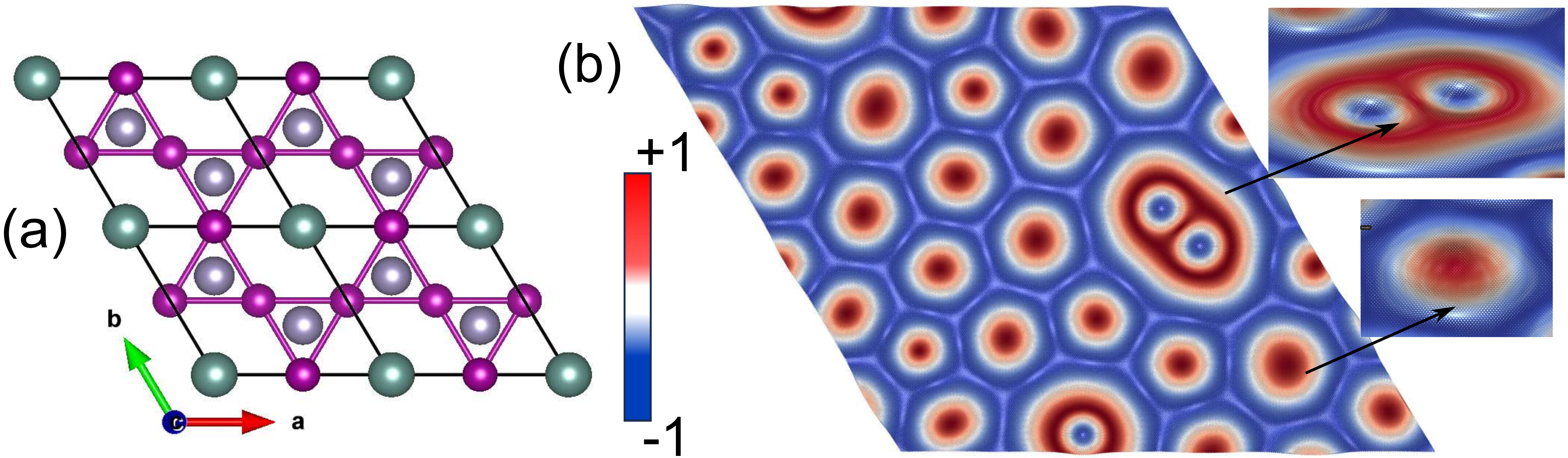}
\caption{(a) Crystal structure of quasi-two-dimensional YMn$_6$Sn$_6$.  The green,  purple and gray atoms represent Y, Mn and Sn atoms respectively.  (b) Planar Dzyaloshinskii–Moriya interaction driven disordered skyrmion state in quasi-2D YMn$_6$Sn$_6$.}
\label{fig2}
\end{figure}

\begin{figure}[ht]
\centering
\includegraphics[width=0.45\textwidth,angle=0]{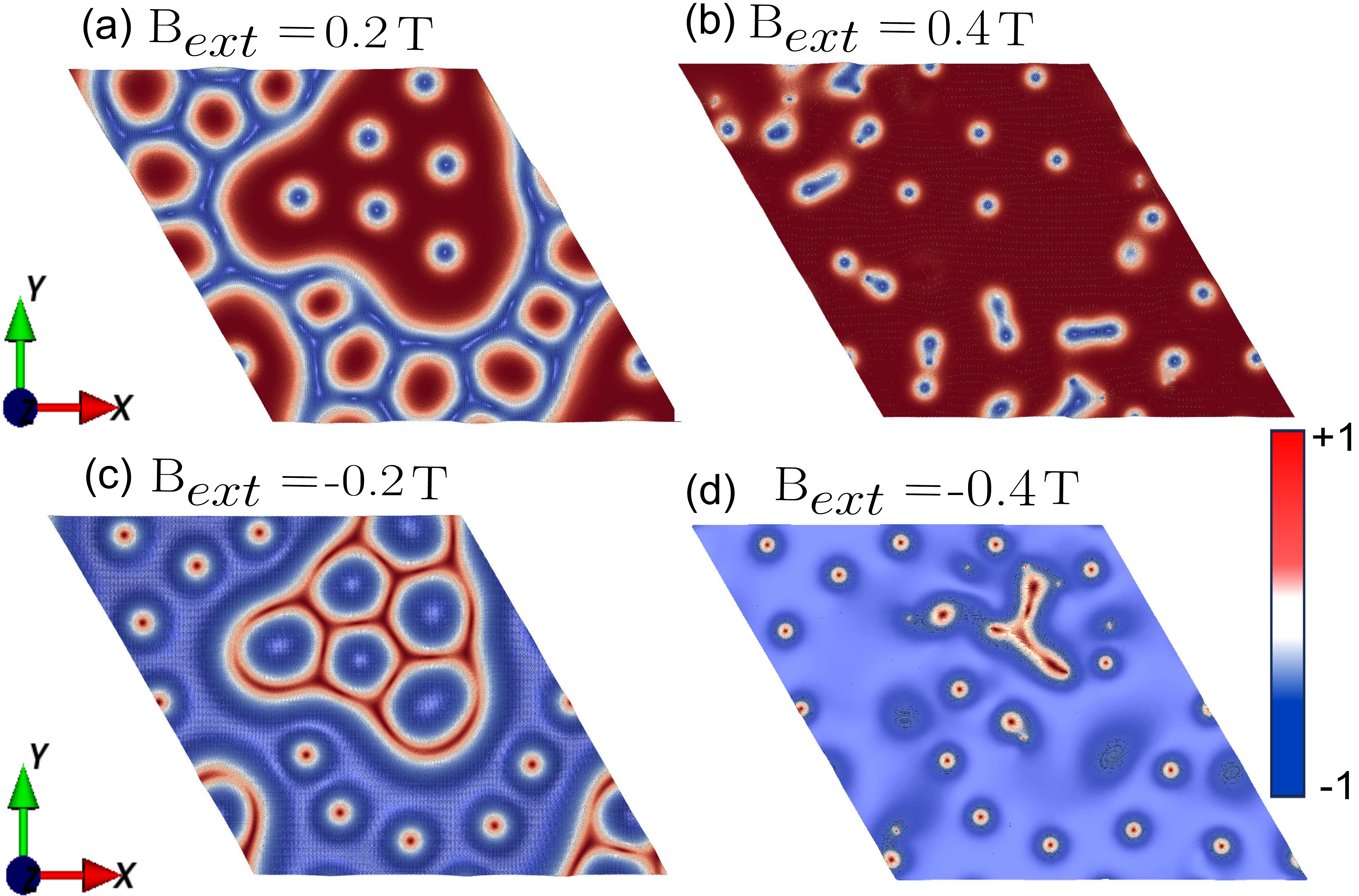}
\caption{Effect of external magnetic field on disordered skyrmion in quasi-two-dimensional structure of YMn$_6$Sn$_6$ with (a) B$_{ext}$ = 0.2 T, (b) B$_{ext}$ = 0.4 T with $B^{\mathrm{ext}}$ = (0,  0,  B$_z$) direction and (c)-(d) with $B^{\mathrm{ext}}$ = (0,  0, -B$_z$) direction respectively. }
\label{fig3}
\end{figure}

\section{Planar DMI  driven disordered skyrmion in YMn$_6$Sn$_6$}
Bulk YMn$_6$Sn$_6$ consists of two inequivalent Mn kagome planes along the $c$-axis, which give rise to two incommensurate spin spirals driven by exchange frustration along $c$-axis \cite{PhysRevB.110.174412}.  The quasi-2D structure of YMn$_6$Sn$_6$ consists of one Mn Kagome plane, as shown in Fig.\ref{fig2} (a) where the calculated magnetic interactions within DFT are mainly contributed from first and second NN in the Mn kagome plane.  The calculated magnetic exchange J$_{ij}$ for first- and second-NN are 5.79 meV and 1.28 meV, respectively, indicating that the Mn kagome planes are ferromagnetic.   On the other hand, the DMI strength D$_{ij}$  for first and second NN are 0.25 meV and 0.46 meV, respectively (see Appendix).  The in-plane components of DMI are found to be  $(D_{x},D_{y})$=(0.22,0.13) meV for first NN, and (0.46,0.07) meV for second NN.  The out-of-plane DMI component D$_{z}$ vanishes in the quasi-2D structure of YMn$_6$Sn$_6$. The calculated magnetic parameters are summarized in Table \ref{tab:tab1}.

\begin{table}[t]
\caption{Calculated exchange interactions $J_{ij}$ and Dzyaloshinskii-Moriya interaction (DMI) components for the first and second nearest-neighbor (NN) Mn pairs in quasi-2D YMn$_6$Sn$_6$. }
\label{tab:tab1}
\begin{ruledtabular}
\begin{tabular}{lccccc}
Neighbor & $J_{ij}$ & $D_x$ & $D_y$ & $D_z$ & $|\mathbf{D}|$ \\
         & (meV)    & (meV) & (meV) & (meV) & (meV) \\
\hline
1st NN & 5.79 & 0.22 & 0.13 & 0 & 0.25 \\
2nd NN & 1.28 & 0.46 & 0.07 & 0 & 0.46 \\
\end{tabular}
\end{ruledtabular}
\end{table}

\par Real-space magnetic configurations are confirmed by a combination of Monte Carlo (MC) annealing and SD simulations (see Appendix) after incorporating the DFT parameters J$_{ij}$,  and D$_{ij}$. The spin Hamiltonian is given by 
\begin{equation}
    {\mathrm{H}}= -\sum_{i,j} {\mathrm{J}}_{ij} {\bm {\mathrm{S}}}_i \cdot {\bm {\mathrm{S}}}_j - \sum_{i,j} {\bm {\mathrm{D}}}_{ij} \cdot ({\bm {\mathrm{S}}}_i \times {\bm {\mathrm{S}}}_j) - \sum_{i} \mu_{i}\bm{B}_{\mathrm{ext}} \cdot \bm{S}_{i} \nonumber \\ \label{eq:spin-Ham}
\end{equation}
The third term is the Zeeman term, where $B_{\mathrm{ext}}$ is the applied external magnetic field and $\mu_{i}$ is the magnetic moment strength at site $i$. The obtained ground state for quasi-2D YMn$_6$Sn$_6$ is a disordered skyrmion phase as presented in Fig.\ref{fig2} (b). The convergence of the total energy and skyrmion number during the MC+SD iterative procedure (see Appendix) confirms the stability of the obtained magnetic ground state. With increasing external magnetic field disordered phase persists up to $B_{\rm ext} \simeq 0.5$~T with decreasing skyrmion radius. After that, it enters into a field-driven ferromagnetic phase upon further increasing $B_{\rm ext}$, see  Fig.\ref{fig3} (a)-(b) and (c)-(d) for positive and negative magnetic field, respectively. The calculated dynamical structure factor (see Appendix) for both disordered and ferromagnetic phases of quasi-2D YMn$_6$Sn$_6$ are presented in Fig.\ref{fig4} (a)-(b) respectively.  To be precise, Fig.\ref{fig4} (a) indicates a clear signature of disordered skyrmion phase in quasi-2D YMn$_6$Sn$_6$ with connected hexagonal rings, supporting the experimental finding \cite{doi:10.1126/sciadv.aar7043}. With increasing the external magnetic field, neighbouring spins are strongly correlated in the ferromagnetic phase with concentric hexagonal rings for $B_{\rm ext} = 0.8$~T,  as shown in Fig.\ref{fig4}(b).

\begin{figure}[ht]
\centering
\includegraphics[width=0.43\textwidth,angle=0]{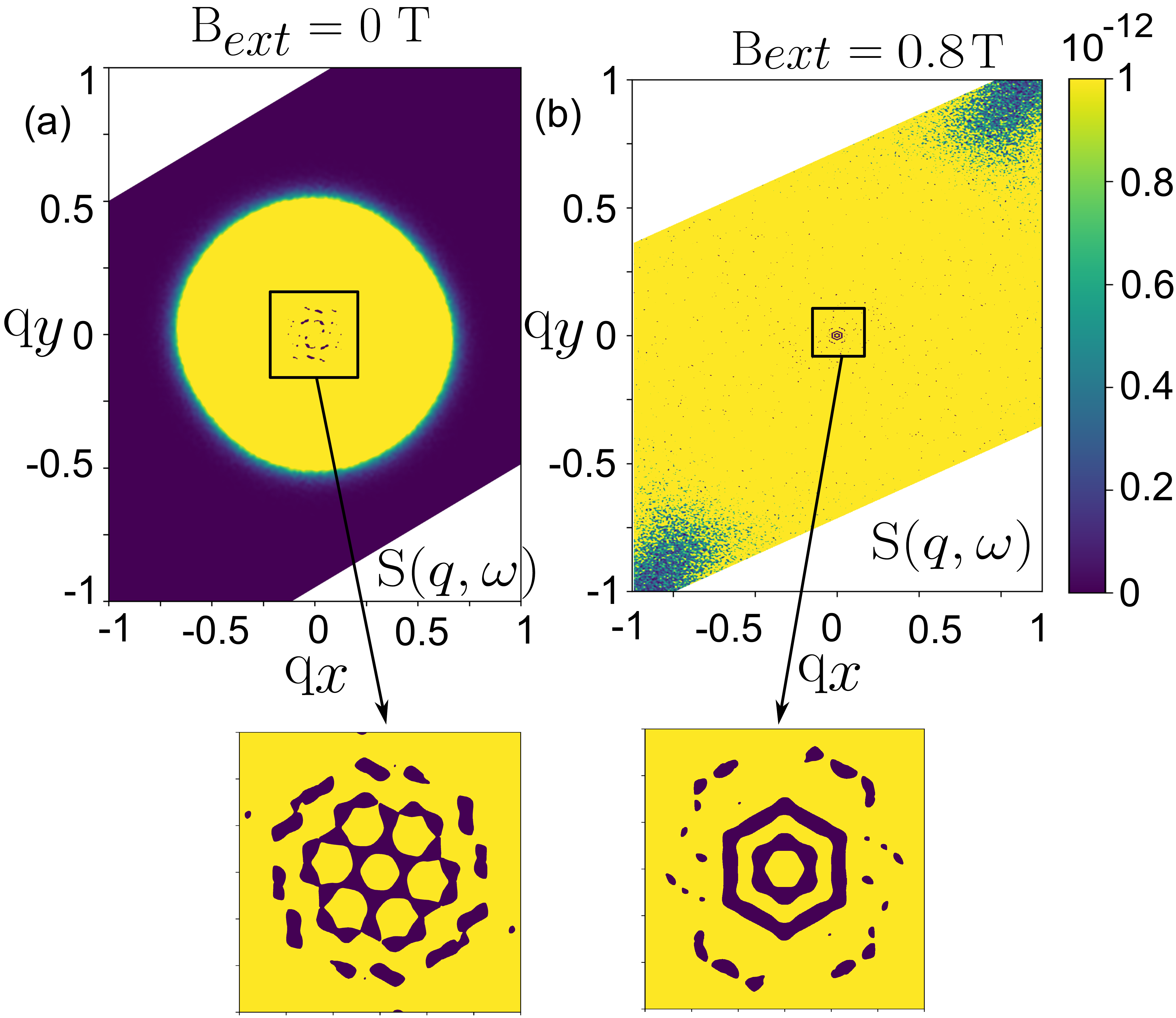}
\caption{Intensity maps of the dynamical structure factor $S_{\text{tot}}(q, \omega)$ in the $q$-plane for constant-$\omega$ scans of (a) disordered skyrmion and (b) ferromagnetic phases in quasi-two-dimensional YMn$_6$Sn$_6$.}
\label{fig4}
\end{figure}

\par In the presence of a dominant planar  DMI, there exists a natural mechanism for stabilizing disordered skyrmionic textures within the kagome lattice.  It leads to a proliferation of locally stable skyrmion-like spin textures that lack long-range order.   This results in a disordered skyrmion phase characterized by finite scalar spin chirality and spatially fluctuating topological charge density.  The disordered state persists up to $B_{\rm ext} \simeq 0.5$~T, with a decreasing skyrmion radius that inhibits long-range lattice formation while retaining the topological character of the spin textures.  Consequently, YMn$_6$Sn$_6$ emerges as a promising platform where planar DMI drives a robust yet disordered skyrmion state,  providing the microscopic origin for unconventional transport responses such as an emergent  THE that we investigate below.

\begin{figure*}[ht]
\centering
\includegraphics[width=1.0\textwidth,angle=0]{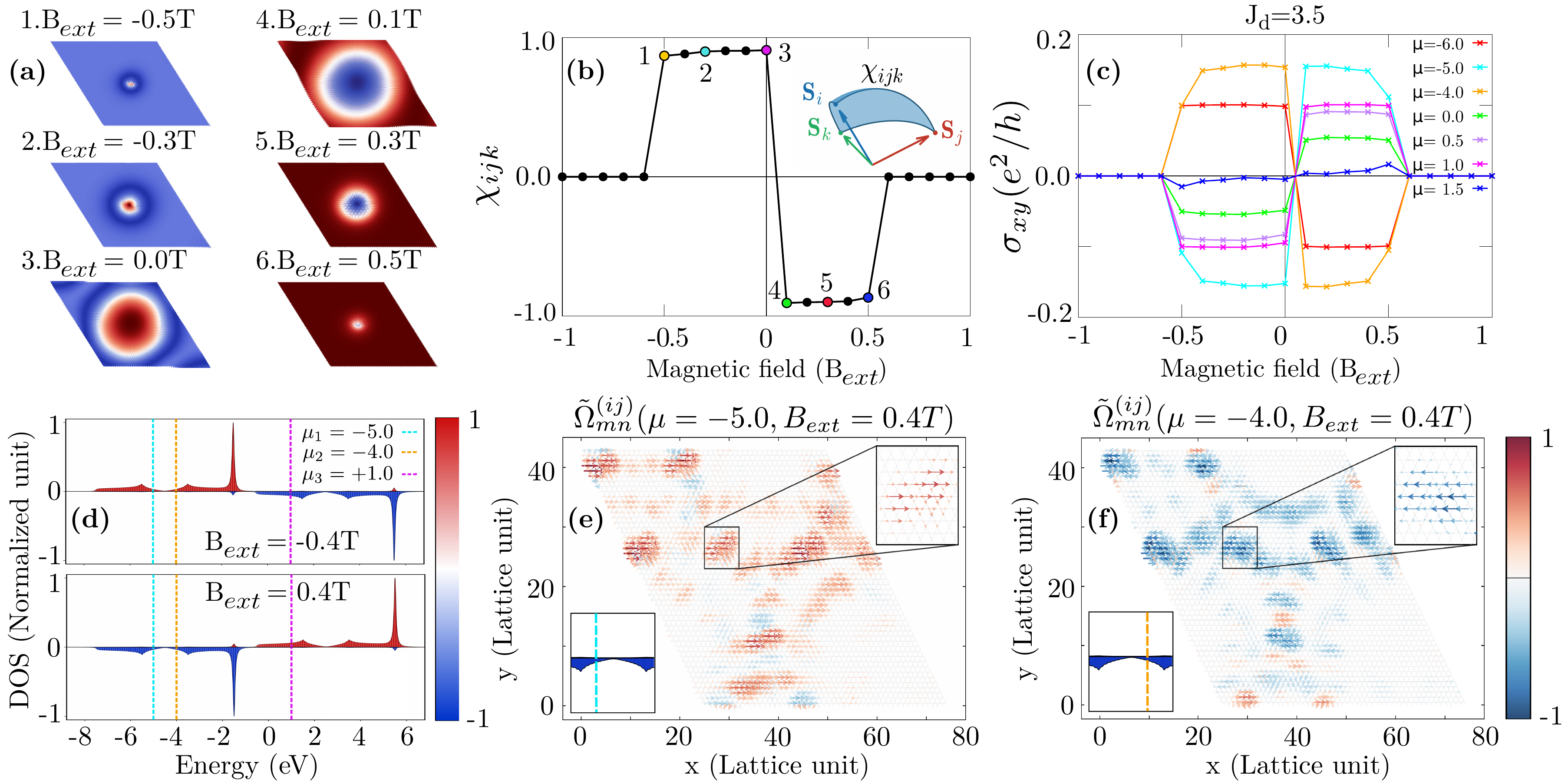}
\caption{(a) Spin textures at six representative $B_\mathrm{ext}$ values.
(b) Scalar spin chirality ($\chi_{ijk}$) as a function of $B_\mathrm{ext}$.
(c) Hall conductivity ($\sigma_{xy}$) versus $B_\mathrm{ext}$ for $J_\mathrm{sd}=3.5$ at different chemical potentials $\mu$.
(d) Spin-resolved density of states (DOS) at $B_\mathrm{ext}=\pm0.4\mathrm{T}$.
(e,f) Spatial distribution of
Berry curvature numerator, computed using the bond-resolved
current operator along x direction,  $\tilde\Omega_{mn}^{(ij)}$ at $\mu = -5.0$ and $\mu = -4.0$ 
    respectively, for $B_\mathrm{ext} = +0.4\,\mathrm{T}$. Insets show the corresponding spin-resolved DOS with $\mu$ marked by a dashed line. $\tilde\Omega_{mn}^{(ij)}$ changes sign, accompanied by a reversal of its flow pattern across the Dirac points.}
\label{fig5}
\end{figure*}

\section{Topological Hall plateau in disordered skyrmion of YMn$_6$Sn$_6$}
We now investigate the topological Hall conductivity using the spin texture of disordered skyrmion obtained from a combination of MC annealing and SD simulations described in the previous section.  We consider a double-exchange Hamiltonian of a kagome lattice which describes the spin coupling to mobile electrons and hosts lattice mapping of quasi-2D  YMn$_6$Sn$_6$ as follows  \cite{PhysRev.100.675}
\begin{align}
{\cal{H}}_{DE}=-t\sum_{\langle ij \rangle} c_{i \sigma}^{\dagger}c_{j \sigma} -J_d \sum_{i,\sigma,\sigma^{\prime}} (\mathbf{S}_i \cdot \boldsymbol{\sigma}_{\sigma \sigma^{\prime}} ) c_{i \sigma}^{\dagger}c_{i \sigma^{\prime}} \nonumber, 
\label{HDE}
\end{align}
where $t$ is the nearest-neighbor electron hopping energy and $J_d$ (Hund's coupling) is the exchange coupling strength between the itinerant electron $\boldsymbol{\sigma}$ and localized spins $\mathbf{S}_i$ of disordered skyrmion generated by a combination of MC annealing and SD simulations (see Appendix).  The Hall conductivity is obtained via the Kubo formula given by \cite{PhysRevB.102.064430}
\begin{align}
\sigma_{xy}\!=\!\frac{e^2}{h}\frac{2\pi}{N} \! \sum_{\epsilon_m \neq \epsilon_n} \! \frac{f_m-f_n}{(\epsilon_m-\epsilon_n)^2+\eta^2} \text{Im}\Big( \langle m | \hat{j_x} | n \rangle \langle n | \hat{j_y} | m \rangle \Big) \nonumber,
\end{align}
where $f_m$ is the Fermi function at temperature $T$, $\epsilon_m$ is the energy for $m^{\rm th}$ eigenstate of ${\cal{H}}_{DE}$, and $\eta$ is the relaxation rate. $\hat{j}_l$=$i\sum_{i\sigma}(t_{l} c_{i\sigma}^{\dagger} c_{i+\hat{l}\sigma}-t_{l}^* c_{i+\hat{l}\sigma}^{\dagger} c_{i\sigma})$ is the current operator along the $l = x,y $ direction, $\sigma= \uparrow, \downarrow$. We used $T \!=\! 0$,  $t \!=\! 1$, $0.5< J_d <4.0$, and broadening parameter $\eta \!=\! 0.1$ eV related to relaxation time in femtosecond scale (10$^{-15}$ s).

\par To establish the robustness of the topological Hall response in quasi-2D YMn$_6$Sn$_6$, we perform large-scale MC annealing and SD simulations on a 300$\times$300$\times$1 lattice size, confirming the disordered skyrmion phase as the ground state.  Owing to the computational cost of direct diagonalization for such system sizes, we adopt a physically transparent coarse-graining strategy. A reduced 50$\times$50$\times$1 lattice containing a single representative skyrmion, see Fig.\ref{fig5} (a), is used to compute the scalar spin chirality $\chi_{ijk}$ and the corresponding topological Hall conductivity $\sigma_{xy}$ at that $B_{\rm ext}$, followed by spatial averaging over the disordered ensembles to recover the statistical response.

\par Note that $\chi_{ijk}$ can be interpreted as a measure of the real-space Berry curvature which remains nearly constant within the external field range  -0.5 $\leq B_{\mathrm{ext}} \leq 0.5$ T in quasi-2D YMn$_6$Sn$_6$, even though the skyrmion radius decreases with increasing $B_{\rm ext}$ as shown in Fig.\ref{fig5}(b). The real space  profile of $\chi_{ijk}$ clearly indicates that while the size of individual skyrmions changes, their topological charge is preserved (see Appendix). As a result, the overall chirality, obtained after spatial averaging over the disordered texture, stays almost unchanged. This robustness shows that the topological properties are not sensitive to the exact size or shape of skyrmions, but are instead governed by the twist profile of the spins carrying the signature of the underlying topology. This ensures the stable emergence of a constant Berry curvature, which is an essential ingredient for a plateau-like profile of topological Hall conductivity in quasi-2D YMn$_6$Sn$_6$.  
This behavior indicates a possible quantization phenomenon as explored below.

\par The anti-symmetric structure i.e.,  $\chi_{ijk}(-B_{\mathrm{ext}})= -\chi_{ijk}(B_{\mathrm{ext}})$ is directly manifested in THE where $\sigma_{xy}(-B_{\mathrm{ext}})=-\sigma_{xy}(B_{\mathrm{ext}})$. The constant profile of topological Hall conductivity within a wide range of the external field  $ |B_{\mathrm{ext}}| \leq 0.5$ T can be considered as a quantized response, as shown in Fig.\ref{fig5} (c) for $J_{\mathrm{d}}=3.5$ meV. To elucidate the role of Hund’s coupling in stabilizing the quantized Hall response, we compute $\sigma_{xy}$ as a function of the coupling strength $J_{\mathrm{d}}$ and chemical potential $\mu$ (see Appendix).  We find that in the moderate coupling regime 2.5 $\leq J_{\mathrm{d}} \leq$ 4 meV, it exhibits a robust, nearly quantized plateau over a range of $\mu$  signaling that the conduction electron spins are tightly aligned with the topological spin  texture. 
In this limit, electrons adiabatically follow the noncollinear spin background, acquiring a well-defined Berry phase that directly reflects the underlying constant scalar spin chirality. This leads to a stable and quantized topological Hall response. Such a plateau-like quantized, and anti-symmetric response is also noticed for orbital and spin Hall effects as well \cite{saha2026quantized}.

\par The anti-symmetric profile of conductivity can be understood from the spin-resolved  density of states (DOS) ranging from negative to positive energy. In the presence of Hund's coupling,  these states  are almost spin-degenerate for  $B_{\mathrm{ext}} = 0$ while up- and down-(down- and up-)spin dominate positive and negative energy states, respectively, for $B_{\mathrm{ext}} > 0$ ($B_{\mathrm{ext}} < 0$) (see Appendix). This field-driven spin-selective reconstruction of the electronic states directly induces a reversal of the quantized Hall conductivity, establishing a clear correlation between spin polarization and topological transport. A representative DOS spectra for $B_{\mathrm{ext}}=\pm 0.4$ T are displayed in Fig.~\ref{fig5}(d).

\par It is important to note that  Hund’s coupling breaks the spin-degeneracy even in the absence of an external magnetic field, leading to spin-resolved DOS while maintaining the kagome profile of DOS for each spin sector.
The distinct flat band and Dirac point can be identified from spin-resolved DOS  and DOS ceases to exist in an intermediate energy window in the strong coupling regime $J_{\mathrm{d}} \gtrsim 4$ meV (see Appendix). This leads to an interesting interplay between Fermi energy and Hund’s coupling as far as the plateau-like profile of $\sigma_{xy}$ is concerned. Importantly, the conductivity plateau changes its value with $\mu$ and $J_{\mathrm{d}}$ while the extent of the plateau as a function of $B_{\mathrm{ext}}$ remains insensitive to the above parameters. This indicates that the quantization of conductivity is not a universal one, but rather relative, existing only with $B_{\mathrm{ext}}$ for a particular set of $(J_{\mathrm{d}},\mu)$ values. We examine below the reason behind such a plateau profile with $\mu$.

\par Interestingly, the response changes its sign and value with $\mu$, which indicates a qualitative dependence on the DOS at the Fermi energy for the conductivity. For $\mu$ values chosen around the Dirac points and van Hove points, $\sigma_{xy}$ exhibits sign reversal at the same fixed value of $B_{\mathrm{ext}}$. The sign inversion originates from the redistribution of Berry curvature numerator near the vanishing dip and local peak of the DOS where the band velocities are finite. Considering the nearest valence and conduction bands, designated by $m$- and $n$-th eigenstates, respectively, around the Fermi energy $\mu$, one can infer important information from the Fermi surface-activated Berry curvature numerator.

\par To explore it, we computed the spatial distribution of Berry curvature numerator using the bond-resolved current operator along x direction and is given by $\tilde\Omega_{mn}^{(ij)}
=
\mathrm{Im}
\left[
\langle m|\hat{J}_x^{\,ij}|n\rangle
\langle n|\hat{J}_y|m\rangle
\right]$ (see Appendix).  The band structure hosts the Dirac points appear at $\mu$ = -4.5 eV. Figure \ref{fig5}(e)-(f) represent $\tilde\Omega_{mn}^{(ij)}$ for $B_{\mathrm{ext}}=+0.4$ T at chemical potential $\mu$ = -5.0 and -4.0 eV respectively which are two different sides of Dirac points respectively. Here the bond contributions are distinct in terms of their magnitude and directions. It exhibits a sign reversal as the chemical potential crosses the Dirac points. Note that bond-resolved profile clearly indicates that only one component of spin participates in the topological Hall transport, while their flow pattern changes on either side of the Dirac and van Hove singularity points \cite{PhysRevB.111.075114,gobel2017signatures} (see Appendix).  The conductivity becomes insignificant when the chemical potential is kept close to flat band. This can be attributed to the vanishingly small group velocity associated with the flat band states. On the other hand, the group velocity experiences a characteristic change around the Dirac point and van Hove singularity which could be manifested through the topological Hall response. However, the response is governed by the anomalous velocity, originating from the spatial distribution of Berry curvature, as nicely captured by the Fermi surface-activated above quantity. Together, these results reveal a highly tunable interplay among spin polarization, Berry curvature, and topology, highlighting the kagome lattice as a versatile platform for controllable topological magnetotransport.

\section{Magnon topology in quasi-2D YMn$_6$Sn$_6$}
 The quasi-2D YMn$_6$Sn$_6$ provides a fertile platform for realizing nontrivial magnon topology that originates from the interplay of exchange interactions between Mn kagome plane and relativistic SOC. We use the spin Hamiltonian Eq. (\ref{eq:spin-Ham}) without magnetic field to produce magnon bands for system. Without SOC, spin Hamiltonian contains only exchange interactions which produce a Dirac point at 31 meV in magnon spectrum. SOC produces non-zero DMI  which lifts magnon band degeneracies and generates finite Berry curvature in momentum space,  giving rise to topologically nontrivial magnon bands characterized by nonzero Chern numbers \cite{fukui2005chern} (see Appendix).  The observed  topological magnon gap is 0.48 meV in the presence of SOC. The calculated Chern numbers for upper, middle and lower magnon bands of quasi-2D YMn$_6$Sn$_6$ are +1, 0 -1, respectively, as shown in Fig.\ref{fig6}(a).

\begin{figure}[ht]
\centering
\includegraphics[width=0.45\textwidth,angle=0]{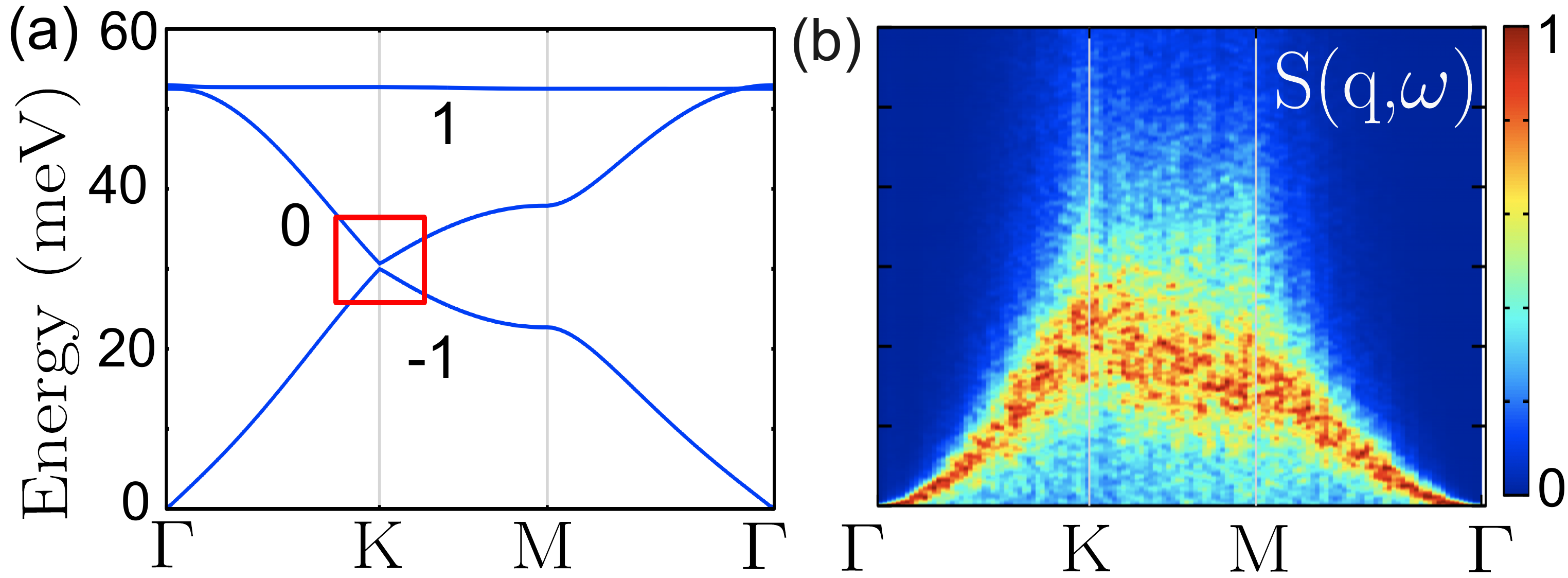}
\caption{(a) Adiabatic magnon spectrum and (b) dynamical structure factor for disordered skyrmion.  The red box indicates topological magnon gap. }
\label{fig6}
\end{figure}

\par The dynamical structure factor associated with the lowest energy magnon band is presented in Fig.\ref{fig6}(b). This represents the disordered skyrmion phase existing in the absence of magnetic field.  The coexistence of disordered skyrmions and topological magnon in  quasi-2D  YMn$_6$Sn$_6$ establishes a unique regime where real-space and momentum-space Berry curvatures are intertwined,  offering a unified framework to understand both topological Hall and magnon Hall effects. These features make the quasi-2D kagome magnets promising candidates for exploring dissipationless spin transport and magnonic devices rooted in topological protection.
 
\section{Experimental Relevance and Significance}
 The DMI-driven disordered skyrmion phase in quasi-2D YMn$_6$Sn$_6$ is experimentally accessible through small-angle neutron scattering  measurements~\cite{doi:10.1126/sciadv.aar7043}, which can directly resolve the emergence of nanoscale chiral spin textures even in the absence of long-range magnetic order. Complementary Lorentz transmission electron microscopy and magnetic force microscopy measurements can further visualize the real-space disordered skyrmion configurations. Importantly, the finite scalar spin chirality associated with these textures generates a measurable topological Hall contribution that can be disentangled from ordinary and anomalous Hall signals through combined longitudinal and transverse magneto-transport measurements, including field- and temperature-dependent Hall resistivity analysis. Remarkably, a topological Hall plateau-like robust response without crystalline skyrmion order demonstrates that a possible topological quantization can survive in a disorder-dominated magnetic regime, extending the conventional paradigm of skyrmion-lattice-driven transport phenomena. This type of Hall signal plateau was recently reported at the Bi$_2$(Se,Te)$_3$/amorphous-EuS interface~\cite{PhysRevB.110.134433}, where such disordered skyrmion states are expected to emerge. These experimentally detectable signatures establish quasi-2D YMn$_6$Sn$_6$ as a promising platform for realizing disorder-resilient topological transport and emergent chirality-driven quantum responses in kagome magnets.

\section{Conclusions}
In conclusion, we establish a material-specific,  {\it{ab initio}} framework to investigate the topological Hall effect in kagome magnets and apply it to quasi-2D YMn$_6$Sn$_6$. We demonstrate that planar DMI stabilizes a robust disordered skyrmion phase, which gives rise to
a plateau-like topological Hall response over a finite magnetic field window.  The presence of nearly uniform scalar spin chirality and associated real-space Berry curvature underpins this plateau where there exists an intricate interplay between Hund's coupling and Fermi energy. The Fermi surface-activated Berry curvature numerator changes non-trivially around the Dirac point and van Hove singularity. Our study further indicates the existence of topological magnon excitations in this disordered phase. Our results provide a unified picture connecting DMI, real-space topology, and transport, and offer concrete, experimentally testable predictions for kagome materials. Additionally,  this work advances numerical methods for simulating general magnetic systems illustrated here through, but not limited to   skyrmion lattices only and therefore provides insight into the intricate interplay between spin textures and electronic transport.

%%%%%%%%%%%%%%%%%%%%%%%%%%%%%%%%%%%%%%%%%%%%%%%%%%%%%%%%%%%%%%%%%%%%%%%%%%%%%%%

\section*{Acknowledgement}

BS  thanks the Prime Minister’s Early
Career Research Grant (PMECRG) of the Anusandhan National Research Foundation Grant No. ANRF/ECRG/2024/005021/PMS. TN thanks the Advanced Research Grant (ARG) from Anusandhan National Research Foundation Grant No. ANRF/ARG/2025/003163/PS.

\bibliography{ymn6sn6}

@ARTICLE{Nagaosa2013-xt,
  title    = "Topological properties and dynamics of magnetic skyrmions",
  author   = "Nagaosa, Naoto and Tokura, Yoshinori",
  journal  = "Nature Nanotechnology",
  volume   =  8,
  number   =  12,
  pages    = "899--911",
  month    =  dec,
  year     =  2013,
  url = {https://doi.org/10.1038/nnano.2013.243}
}

@article{PhysRevB.103.144308,
  title = {Role of time reversal symmetry and tilting in circular photogalvanic responses},
  author = {Sadhukhan, Banasree and Nag, Tanay},
  journal = {Phys. Rev. B},
  volume = {103},
  issue = {14},
  pages = {144308},
  numpages = {10},
  year = {2021},
  month = {Apr},
  publisher = {American Physical Society},
  doi = {10.1103/PhysRevB.103.144308},
  url = {https://link.aps.org/doi/10.1103/PhysRevB.103.144308}
}

@article{PhysRevB.106.045424,
  title = {Understanding the three-dimensional quantum Hall effect in generic multi-Weyl semimetals},
  author = {Xiong, Feng and Honerkamp, Carsten and Kennes, Dante M. and Nag, Tanay},
  journal = {Phys. Rev. B},
  volume = {106},
  issue = {4},
  pages = {045424},
  numpages = {18},
  year = {2022},
  month = {Jul},
  publisher = {American Physical Society},
  doi = {10.1103/PhysRevB.106.045424},
  url = {https://link.aps.org/doi/10.1103/PhysRevB.106.045424}
}

@article{PhysRevB.104.245122,
  title = {Electronic structure and unconventional nonlinear response in double Weyl semimetal {SrSi$_2$}},
  author = {Sadhukhan, Banasree and Nag, Tanay},
  journal = {Phys. Rev. B},
  volume = {104},
  issue = {24},
  pages = {245122},
  numpages = {10},
  year = {2021},
  month = {Dec},
  publisher = {American Physical Society},
  doi = {10.1103/PhysRevB.104.245122},
  url = {https://link.aps.org/doi/10.1103/PhysRevB.104.245122}
}

@article{PhysRevB.105.214307,
  title = {Distinct signatures of particle-hole symmetry breaking in transport coefficients for generic multi-Weyl semimetals},
  author = {Nag, Tanay and Kennes, Dante M.},
  journal = {Phys. Rev. B},
  volume = {105},
  issue = {21},
  pages = {214307},
  numpages = {16},
  year = {2022},
  month = {Jun},
  publisher = {American Physical Society},
  doi = {10.1103/PhysRevB.105.214307},
  url = {https://link.aps.org/doi/10.1103/PhysRevB.105.214307}
}

@article{PhysRevB.110.134433,
  title = {Emergence of planar topological Hall anisotropy in {Bi$_2$(Se,Te)$_3$} by proximity-induced spin-canted state of the Heisenberg ferromagnetic insulator {EuS}},
  author = {Suri, Dhavala and Sasmal, Satyaki and Bhardwaj, Archit and Singh, Juhi and Mundlia, Suman and Mishra, Anshika and Mohanta, Narayan and Raman, Karthik V.},
  journal = {Phys. Rev. B},
  volume = {110},
  issue = {13},
  pages = {134433},
  numpages = {7},
  year = {2024},
  month = {Oct},
  publisher = {American Physical Society},
  doi = {10.1103/PhysRevB.110.134433},
  url = {https://link.aps.org/doi/10.1103/PhysRevB.110.134433}
}

@article{saha2026quantized,
  title={Quantized orbital and spin Hall transport: interplay between $ sp $-hybridization, altermagnetism and spin-orbit coupling},
  author={Saha, Saikat and Sadhukhan, Banasree and Nag, Tanay},
  journal={arXiv preprint arXiv:2606.01404},
  year={2026}
}

@article{PhysRevB.107.245141,
  title = {Third-order Hall effect in the surface states of a topological insulator},
  author = {Nag, Tanay and Das, Sanjib Kumar and Zeng, Chuanchang and Nandy, Snehasish},
  journal = {Phys. Rev. B},
  volume = {107},
  issue = {24},
  pages = {245141},
  numpages = {8},
  year = {2023},
  month = {Jun},
  publisher = {American Physical Society},
  doi = {10.1103/PhysRevB.107.245141},
  url = {https://link.aps.org/doi/10.1103/PhysRevB.107.245141}
}

@article{PhysRevB.106.045417,
  title = {Correlated disorder induced anomalous transport in magnetically doped topological insulators},
  author = {Okugawa, Takuya and Nag, Tanay and Kennes, Dante M.},
  journal = {Phys. Rev. B},
  volume = {106},
  issue = {4},
  pages = {045417},
  numpages = {7},
  year = {2022},
  month = {Jul},
  publisher = {American Physical Society},
  doi = {10.1103/PhysRevB.106.045417},
  url = {https://link.aps.org/doi/10.1103/PhysRevB.106.045417}
}

@article{PhysRevB.104.115420,
  title = {Topological Magnus responses in two- and three-dimensional systems},
  author = {Das, Sanjib Kumar and Nag, Tanay and Nandy, Snehasish},
  journal = {Phys. Rev. B},
  volume = {104},
  issue = {11},
  pages = {115420},
  numpages = {15},
  year = {2021},
  month = {Sep},
  publisher = {American Physical Society},
  doi = {10.1103/PhysRevB.104.115420},
  url = {https://link.aps.org/doi/10.1103/PhysRevB.104.115420}
}

@article{Nag_2025,
doi = {10.1088/1361-648X/adb6e9},
url = {https://doi.org/10.1088/1361-648X/adb6e9},
year = {2025},
month = {feb},
publisher = {IOP Publishing},
volume = {37},
number = {15},
pages = {153001},
author = {Nag, Tanay and Mandal, Saptarshi},
title = {Extended Haldane model- a modern gateway to topological insulators},
journal = {Journal of Physics: Condensed Matter}
}

@article{PhysRevB.103.235154,
  title = {Eightfold quantum Hall phases in a time reversal symmetry broken tight binding model},
  author = {Saha, Sudarshan and Nag, Tanay and Mandal, Saptarshi},
  journal = {Phys. Rev. B},
  volume = {103},
  issue = {23},
  pages = {235154},
  numpages = {7},
  year = {2021},
  month = {Jun},
  publisher = {American Physical Society},
  doi = {10.1103/PhysRevB.103.235154},
  url = {https://link.aps.org/doi/10.1103/PhysRevB.103.235154}
}

@article{Zhang_2023,
doi = {10.1088/2752-5724/ace1df},
url = {https://doi.org/10.1088/2752-5724/ace1df},
year = {2023},
month = {jul},
publisher = {IOP Publishing},
volume = {2},
number = {3},
pages = {032201},
author = {Zhang, Huai and Zhang, Yajiu and Hou, Zhipeng and Qin, Minghui and Gao, Xingsen and Liu, Junming},
title = {Magnetic skyrmions: materials, manipulation, detection, and applications in spintronic devices},
journal = {Materials Futures}
}

@ARTICLE{Rosler2006-xa,
  title    = "Spontaneous skyrmion ground states in magnetic metals",
  author   = "R{\"o}{\ss}ler, U K and Bogdanov, A N and Pfleiderer, C",
  journal  = "Nature",
  volume   =  442,
  number   =  7104,
  pages    = "797--801",
  month    =  aug,
  year     =  2006,
  url = {https://doi.org/10.1038/nature05056}
}

@article{doi:10.1126/science.1166767,
author = {S. Mühlbauer  and B. Binz  and F. Jonietz  and C. Pfleiderer  and A. Rosch  and A. Neubauer  and R. Georgii  and P. Böni },
title = {Skyrmion Lattice in a Chiral Magnet},
journal = {Science},
volume = {323},
number = {5916},
pages = {915-919},
year = {2009},
doi = {10.1126/science.1166767},
URL = {https://www.science.org/doi/abs/10.1126/science.1166767},
eprint = {https://www.science.org/doi/pdf/10.1126/science.1166767}
}

@ARTICLE{Yu2010-ys,
  title    = "Real-space observation of a two-dimensional skyrmion crystal",
  author   = "Yu, X Z and Onose, Y and Kanazawa, N and Park, J H and Han, J H
              and Matsui, Y and Nagaosa, N and Tokura, Y",
  journal  = "Nature",
  volume   =  465,
  number   =  7300,
  pages    = "901--904",
  month    =  jun,
  year     =  2010,
    url = {https://doi.org/10.1038/nature09124}
}

@ARTICLE{Heinze2011-wt,
  title    = "Spontaneous atomic-scale magnetic skyrmion lattice in two
              dimensions",
  author   = "Heinze, Stefan and von Bergmann, Kirsten and Menzel, Matthias and
              Brede, Jens and Kubetzka, Andr{\'e} and Wiesendanger, Roland and
              Bihlmayer, Gustav and Bl{\"u}gel, Stefan",
  journal  = "Nature Physics",
  volume   =  7,
  number   =  9,
  pages    = "713--718",
  month    =  sep,
  year     =  2011,
      url = {https://doi.org/10.1038/nphys2045}
}

@ARTICLE{Khanh2020-of,
  title    = "Nanometric square skyrmion lattice in a centrosymmetric
              tetragonal magnet",
  author   = "Khanh, Nguyen Duy and Nakajima, Taro and Yu, Xiuzhen and Gao,
              Shang and Shibata, Kiyou and Hirschberger, Max and Yamasaki,
              Yuichi and Sagayama, Hajime and Nakao, Hironori and Peng, Licong
              and Nakajima, Kiyomi and Takagi, Rina and Arima, Taka-Hisa and
              Tokura, Yoshinori and Seki, Shinichiro",
  journal  = "Nature Nanotechnology",
  volume   =  15,
  number   =  6,
  pages    = "444--449",
  month    =  jun,
  year     =  2020,
  url = {https://doi.org/10.1038/s41565-020-0684-7} 
}

@article{PhysRevLett.115.036602,
  title = {Dzyaloshinskii-Moriya Interaction and Hall Effects in the Skyrmion Phase of {Mn$_{1-x}$Fe$_x$Ge}},
  author = {Gayles, J. and Freimuth, F. and Schena, T. and Lani, G. and Mavropoulos, P. and Duine, R. A. and Bl\"ugel, S. and Sinova, J. and Mokrousov, Y.},
  journal = {Phys. Rev. Lett.},
  volume = {115},
  issue = {3},
  pages = {036602},
  numpages = {6},
  year = {2015},
  month = {Jul},
  publisher = {American Physical Society},
  doi = {10.1103/PhysRevLett.115.036602},
  url = {https://link.aps.org/doi/10.1103/PhysRevLett.115.036602}
  }

@article{PhysRevLett.116.247201,
  title = {Dzyaloshinskii-Moriya Interaction as a Consequence of a Doppler Shift due to Spin-Orbit-Induced Intrinsic Spin Current},
  author = {Kikuchi, Toru and Koretsune, Takashi and Arita, Ryotaro and Tatara, Gen},
  journal = {Phys. Rev. Lett.},
  volume = {116},
  issue = {24},
  pages = {247201},
  numpages = {6},
  year = {2016},
  month = {Jun},
  publisher = {American Physical Society},
  doi = {10.1103/PhysRevLett.116.247201},
  url = {https://link.aps.org/doi/10.1103/PhysRevLett.116.247201}
}

@article{PhysRevLett.125.117204,
  title = {Formation Mechanism of the Helical Q Structure in Gd-Based Skyrmion Materials},
  author = {Nomoto, Takuya and Koretsune, Takashi and Arita, Ryotaro},
  journal = {Phys. Rev. Lett.},
  volume = {125},
  issue = {11},
  pages = {117204},
  numpages = {6},
  year = {2020},
  month = {Sep},
  publisher = {American Physical Society},
  doi = {10.1103/PhysRevLett.125.117204},
  url = {https://link.aps.org/doi/10.1103/PhysRevLett.125.117204}
  }

@article{PhysRevLett.102.186602,
  title = {Topological Hall Effect in the $A$ Phase of {MnSi}},
  author = {Neubauer, A. and Pfleiderer, C. and Binz, B. and Rosch, A. and Ritz, R. and Niklowitz, P. G. and B\"oni, P.},
  journal = {Phys. Rev. Lett.},
  volume = {102},
  issue = {18},
  pages = {186602},
  numpages = {4},
  year = {2009},
  month = {May},
  publisher = {American Physical Society},
  doi = {10.1103/PhysRevLett.102.186602},
  url = {https://link.aps.org/doi/10.1103/PhysRevLett.102.186602}
}

@article{PhysRevB.99.174425,
  title = {Weak coupling theory of topological Hall effect},
  author = {Nakazawa, Kazuki and Kohno, Hiroshi},
  journal = {Phys. Rev. B},
  volume = {99},
  issue = {17},
  pages = {174425},
  numpages = {28},
  year = {2019},
  month = {May},
  publisher = {American Physical Society},
  doi = {10.1103/PhysRevB.99.174425},
  url = {https://link.aps.org/doi/10.1103/PhysRevB.99.174425}
}

@article{PhysRevLett.117.027202,
  title = {Electron Scattering on a Magnetic Skyrmion in the Nonadiabatic Approximation},
  author = {Denisov, K. S. and Rozhansky, I. V. and Averkiev, N. S. and L\"ahderanta, E.},
  journal = {Phys. Rev. Lett.},
  volume = {117},
  issue = {2},
  pages = {027202},
  numpages = {5},
  year = {2016},
  month = {Jul},
  publisher = {American Physical Society},
  doi = {10.1103/PhysRevLett.117.027202},
  url = {https://link.aps.org/doi/10.1103/PhysRevLett.117.027202}
}

@article{PhysRevB.98.195439,
  title = {General theory of the topological Hall effect in systems with chiral spin textures},
  author = {Denisov, K. S. and Rozhansky, I. V. and Averkiev, N. S. and L\"ahderanta, E.},
  journal = {Phys. Rev. B},
  volume = {98},
  issue = {19},
  pages = {195439},
  numpages = {12},
  year = {2018},
  month = {Nov},
  publisher = {American Physical Society},
  doi = {10.1103/PhysRevB.98.195439},
  url = {https://link.aps.org/doi/10.1103/PhysRevB.98.195439}
}

@article{PhysRevB.92.115417,
  title = {Quantized topological Hall effect in skyrmion crystal},
  author = {Hamamoto, Keita and Ezawa, Motohiko and Nagaosa, Naoto},
  journal = {Phys. Rev. B},
  volume = {92},
  issue = {11},
  pages = {115417},
  numpages = {6},
  year = {2015},
  month = {Sep},
  publisher = {American Physical Society},
  doi = {10.1103/PhysRevB.92.115417},
  url = {https://link.aps.org/doi/10.1103/PhysRevB.92.115417}
}

@article{PhysRevB.104.174432,
  title = {Skyrmion-size dependence of the topological Hall effect: A real-space calculation},
  author = {Matsui, Akira and Nomoto, Takuya and Arita, Ryotaro},
  journal = {Phys. Rev. B},
  volume = {104},
  issue = {17},
  pages = {174432},
  numpages = {6},
  year = {2021},
  month = {Nov},
  publisher = {American Physical Society},
  doi = {10.1103/PhysRevB.104.174432},
  url = {https://link.aps.org/doi/10.1103/PhysRevB.104.174432}
}

@article{PhysRevB.100.064429,
  title = {Topological Hall effect and emergent skyrmion crystal at manganite-iridate oxide interfaces},
  author = {Mohanta, Narayan and Dagotto, Elbio and Okamoto, Satoshi},
  journal = {Phys. Rev. B},
  volume = {100},
  issue = {6},
  pages = {064429},
  numpages = {9},
  year = {2019},
  month = {Aug},
  publisher = {American Physical Society},
  doi = {10.1103/PhysRevB.100.064429},
  url = {https://link.aps.org/doi/10.1103/PhysRevB.100.064429}
}

@article{PhysRevB.102.064430,
  title = {Planar topological Hall effect from conical spin spirals},
  author = {Mohanta, Narayan and Okamoto, Satoshi and Dagotto, Elbio},
  journal = {Phys. Rev. B},
  volume = {102},
  issue = {6},
  pages = {064430},
  numpages = {6},
  year = {2020},
  month = {Aug},
  publisher = {American Physical Society},
  doi = {10.1103/PhysRevB.102.064430},
  url = {https://link.aps.org/doi/10.1103/PhysRevB.102.064430}
}

@article{PhysRevB.111.134433,
  title = {Topological Hall-like behavior of multidomain ferromagnets},
  author = {Sabri, Houssam and Carlson, Benjamin E. and Pershoguba, Sergey S. and Zang, Jiadong},
  journal = {Phys. Rev. B},
  volume = {111},
  issue = {13},
  pages = {134433},
  numpages = {9},
  year = {2025},
  month = {Apr},
  publisher = {American Physical Society},
  doi = {10.1103/PhysRevB.111.134433},
  url = {https://link.aps.org/doi/10.1103/PhysRevB.111.134433}
}

@article{PhysRevX.15.011054,
  title = {Topological Hall Effect of Skyrmions from first Principles},
  author = {Chen, Hsiao-Yi and Nomoto, Takuya and Hirschberger, Max and Arita, Ryotaro},
  journal = {Phys. Rev. X},
  volume = {15},
  issue = {1},
  pages = {011054},
  numpages = {16},
  year = {2025},
  month = {Mar},
  publisher = {American Physical Society},
  doi = {10.1103/PhysRevX.15.011054},
  url = {https://link.aps.org/doi/10.1103/PhysRevX.15.011054}
}

@article{PhysRevLett.106.156603,
  title = {Large Topological Hall Effect in a Short-Period Helimagnet MnGe},
  author = {Kanazawa, N. and Onose, Y. and Arima, T. and Okuyama, D. and Ohoyama, K. and Wakimoto, S. and Kakurai, K. and Ishiwata, S. and Tokura, Y.},
  journal = {Phys. Rev. Lett.},
  volume = {106},
  issue = {15},
  pages = {156603},
  numpages = {4},
  year = {2011},
  month = {Apr},
  publisher = {American Physical Society},
  doi = {10.1103/PhysRevLett.106.156603},
  url = {https://link.aps.org/doi/10.1103/PhysRevLett.106.156603}
}

@article{PhysRevB.80.054416,
  title = {Skyrmions and anomalous Hall effect in a Dzyaloshinskii-Moriya spiral magnet},
  author = {Yi, Su Do and Onoda, Shigeki and Nagaosa, Naoto and Han, Jung Hoon},
  journal = {Phys. Rev. B},
  volume = {80},
  issue = {5},
  pages = {054416},
  numpages = {6},
  year = {2009},
  month = {Aug},
  publisher = {American Physical Society},
  doi = {10.1103/PhysRevB.80.054416},
  url = {https://link.aps.org/doi/10.1103/PhysRevB.80.054416}
}

@article{doi:10.1126/sciadv.aap9962,
author = {Hiroaki Ishizuka  and Naoto Nagaosa },
title = {Spin chirality induced skew scattering and anomalous Hall effect in chiral magnets},
journal = {Science Advances},
volume = {4},
number = {2},
pages = {eaap9962},
year = {2018},
doi = {10.1126/sciadv.aap9962},
URL = {https://www.science.org/doi/abs/10.1126/sciadv.aap9962},
eprint = {https://www.science.org/doi/pdf/10.1126/sciadv.aap9962}}

@article{doi:10.1126/sciadv.1600304,
author = {Jobu Matsuno  and Naoki Ogawa  and Kenji Yasuda  and Fumitaka Kagawa  and Wataru Koshibae  and Naoto Nagaosa  and Yoshinori Tokura  and Masashi Kawasaki },
title = {Interface-driven topological Hall effect in {SrRuO$_3$-SrIrO$_3$} bilayer},
journal = {Science Advances},
volume = {2},
number = {7},
pages = {e1600304},
year = {2016},
doi = {10.1126/sciadv.1600304},
URL = {https://www.science.org/doi/abs/10.1126/sciadv.1600304},
eprint = {https://www.science.org/doi/pdf/10.1126/sciadv.1600304}}

@ARTICLE{Banerjee2013-lm,
  title    = "Ferromagnetic exchange, spin--orbit coupling and spiral magnetism
              at the {LaAlO$_3$/SrTiO$_3$} interface",
  author   = "Banerjee, Sumilan and Erten, Onur and Randeria, Mohit",
  journal  = "Nature Physics",
  volume   =  9,
  number   =  10,
  pages    = "626--630",
  month    =  oct,
  year     =  2013,
   url = {https://doi.org/10.1038/nphys2702} 
}

@article{PhysRevB.99.104402,
  title = {Quantifying chiral exchange interaction for N\'eel-type skyrmions via Lorentz transmission electron microscopy},
  author = {Jiang, Wanjun and Zhang, Sheng and Wang, Xiao and Phatak, Charudatta and Wang, Qiang and Zhang, Wei and Jungfleisch, Matthias Benjamin and Pearson, John E. and Liu, Yizhou and Zang, Jiadong and Cheng, Xuemei and Petford-Long, Amanda and Hoffmann, Axel and te Velthuis, Suzanne G. E.},
  journal = {Phys. Rev. B},
  volume = {99},
  issue = {10},
  pages = {104402},
  numpages = {9},
  year = {2019},
  month = {Mar},
  publisher = {American Physical Society},
  doi = {10.1103/PhysRevB.99.104402},
  url = {https://link.aps.org/doi/10.1103/PhysRevB.99.104402}
}

@article{doi:10.1126/sciadv.abq2765,
author = {Nishchhal Verma  and Zachariah Addison  and Mohit Randeria },
title = {Unified theory of the anomalous and topological Hall effects with phase-space Berry curvatures},
journal = {Science Advances},
volume = {8},
number = {45},
pages = {eabq2765},
year = {2022},
doi = {10.1126/sciadv.abq2765},
URL = {https://www.science.org/doi/abs/10.1126/sciadv.abq2765},
eprint = {https://www.science.org/doi/pdf/10.1126/sciadv.abq2765}}

@article{Dzyaloshinskii1957,
  author  = {I. Dzyaloshinskii},
  title   = {A thermodynamic theory of “weak” ferromagnetism of antiferromagnetics},
  journal = {Soviet Physics JETP},
  volume  = {5},
  pages   = {1259--1272},
  year    = {1957},
  doi     = {10.1007/BF01326866}
}

@article{Dzyaloshinskii1958,
  author  = {I. Dzyaloshinskii},
  title   = {A thermodynamic theory of weak ferromagnetism of antiferromagnetics},
  journal = {Journal of Physics and Chemistry of Solids},
  volume  = {4},
  pages   = {241--255},
  year    = {1958},
  doi     = {10.1016/0022-3697(58)90076-3}
}

@article{Moriya1960PRL,
  author  = {T. Moriya},
  title   = {New Mechanism of Anisotropic Superexchange Interaction},
  journal = {Physical Review Letters},
  volume  = {4},
  pages   = {228--230},
  year    = {1960},
  doi     = {10.1103/PhysRevLett.4.228}
}

@article{Moriya1960PR,
  author  = {T. Moriya},
  title   = {Anisotropic Superexchange Interaction and Weak Ferromagnetism},
  journal = {Physical Review},
  volume  = {120},
  pages   = {91--98},
  year    = {1960},
  doi     = {10.1103/PhysRev.120.91}
}

@article{Nagaosa2013,
  author  = {N. Nagaosa and Y. Tokura},
  title   = {Topological properties and dynamics of magnetic skyrmions},
  journal = {Nature Nanotechnology},
  volume  = {8},
  pages   = {899--911},
  year    = {2013},
  doi     = {10.1038/nnano.2013.243}
}

@article{doi:10.1126/sciadv.aar7043,
author = {Kosuke Karube  and Jonathan S. White  and Daisuke Morikawa  and Charles D. Dewhurst  and Robert Cubitt  and Akiko Kikkawa  and Xiuzhen Yu  and Yusuke Tokunaga  and Taka-hisa Arima  and Henrik M. Rønnow  and Yoshinori Tokura  and Yasujiro Taguchi },
title = {Disordered skyrmion phase stabilized by magnetic frustration in a chiral magnet},
journal = {Science Advances},
volume = {4},
number = {9},
pages = {eaar7043},
year = {2018},
doi = {10.1126/sciadv.aar7043},
URL = {https://www.science.org/doi/abs/10.1126/sciadv.aar7043},
eprint = {https://www.science.org/doi/pdf/10.1126/sciadv.aar7043}}

@article{fert2017magnetic,
  title = {Magnetic skyrmions: advances in physics and potential applications},
  author = {Fert, Albert and Reyren, Nicolas and Cros, Vincent},
  journal = {Nature Reviews Materials},
  volume = {2},
  pages = {17031},
  year = {2017},
  doi = {10.1038/natrevmats.2017.31}
}

@article{ezawa2011compact,
  title = {Compact merons and skyrmions in thin chiral magnetic films},
  author = {Ezawa, Motohiko},
  journal = {Physical Review B},
  volume = {83},
  number = {10},
  pages = {100408},
  year = {2011},
  doi = {10.1103/PhysRevB.83.100408}
}

@article{heinze2011spontaneous,
  title = {Spontaneous atomic-scale magnetic skyrmion lattice in two dimensions},
  author = {Heinze, Stefan and von Bergmann, Kirsten and Menzel, Matthias and Brede, Jörg and Kubetzka, Andreas and Wiesendanger, Roland and Bihlmayer, Gustav and Blugel, Stefan},
  journal = {Nature Physics},
  volume = {7},
  pages = {713--718},
  year = {2011},
  doi = {10.1038/nphys2045}
}

@article{moreau2016additive,
  title = {Additive interfacial chiral interaction in multilayers for stabilization of small individual skyrmions at room temperature},
  author = {Moreau-Luchaire, C. and Moutafis, C. and Reyren, N. and Sampaio, J. and Vaz, C. A. F. and Van Horne, N. and Bouzehouane, K. and Garcia, K. and Deranlot, C. and Warnicke, P. and others},
  journal = {Nature Nanotechnology},
  volume = {11},
  pages = {444--448},
  year = {2016},
  doi = {10.1038/nnano.2015.313}
}

@article{boulle2016room,
  title = {Room-temperature chiral magnetic skyrmions in ultrathin magnetic nanostructures},
  author = {Boulle, Olivier and Vogel, Julien and Yang, H. and Pizzini, S. and de Souza Chaves, D. and Locatelli, A. and Mentes, T. O. and Sala, A. and Buda-Prejbeanu, L. D. and Klein, O. and others},
  journal = {Nature Nanotechnology},
  volume = {11},
  pages = {449--454},
  year = {2016},
  doi = {10.1038/nnano.2015.315}
}

@article{woo2016observation,
  title = {Observation of room-temperature magnetic skyrmions and their current-driven dynamics in ultrathin metallic ferromagnets},
  author = {Woo, Seonghoon and Litzius, Kai and Kruger, Benjamin and Im, Mi-Young and Caretta, Lorenzo and Richter, Klaus and Mann, Martin and Krone, Arne and Reeve, Robert M. and Weigand, Markus and others},
  journal = {Nature Materials},
  volume = {15},
  pages = {501--506},
  year = {2016},
  doi = {10.1038/nmat4593}
}

@article{PhysRev.100.675,
  title = {Considerations on Double Exchange},
  author = {Anderson, P. W. and Hasegawa, H.},
  journal = {Phys. Rev.},
  volume = {100},
  issue = {2},
  pages = {675--681},
  numpages = {0},
  year = {1955},
  month = {Oct},
  publisher = {American Physical Society},
  doi = {10.1103/PhysRev.100.675},
  url = {https://link.aps.org/doi/10.1103/PhysRev.100.675}
}

@article{SECCHI201561,
title = {Magnetic interactions in strongly correlated systems: Spin and orbital contributions},
journal = {Annals of Physics},
volume = {360},
pages = {61-97},
year = {2015},
issn = {0003-4916},
doi = {https://doi.org/10.1016/j.aop.2015.05.002},
url = {https://www.sciencedirect.com/science/article/pii/S0003491615001864},
author = {A. Secchi and A.I. Lichtenstein and M.I. Katsnelson}
}

@article{PhysRevB.111.075114,
  title = {Higher-order Van Hove singularities in kagome topological bands},
  author = {Wang, Edrick and Pullasseri, Lakshmi and Santos, Luiz H.},
  journal = {Phys. Rev. B},
  volume = {111},
  issue = {7},
  pages = {075114},
  numpages = {10},
  year = {2025},
  month = {Feb},
  publisher = {American Physical Society},
  doi = {10.1103/PhysRevB.111.075114},
  url = {https://link.aps.org/doi/10.1103/PhysRevB.111.075114}
}

@article{gobel2017signatures,
  title={Signatures of lattice geometry in quantum and topological Hall effect},
  author={G{\"o}bel, B{\"o}rge and Mook, Alexander and Henk, J{\"u}rgen and Mertig, Ingrid},
  journal={New Journal of Physics},
  volume={19},
  number={6},
  pages={063042},
  year={2017},
  publisher={IOP Publishing},
  DOI={10.1088/1367-2630/aa709b},
  url = {https://iopscience.iop.org/article/10.1088/1367-2630/aa709b#njpaa709bs4}
}

@article{Ding2006,
  author = {Ding, Hang and Shu, Chang},
  title = {A stencil adaptive algorithm for finite difference solution of incompressible viscous flows},
  journal = {Journal of Computational Physics},
  volume = {214},
  pages = {397-420},
  year = {2006},
  doi = {10.1016/j.jcp.2005.09.021}
}

@book{Eriksson2017,
  author = {Eriksson, Olle and Bergman, Anders and Bergqvist, Lars and Hellsvik, Johan},
  title = {Atomistic Spin Dynamics: Foundations and Applications},
  publisher = {Oxford University Press},
  year = {2017},
  isbn = {9780198788669}
}

@book{wills2010full,
  title={Full-Potential Electronic Structure Method: energy and force calculations with density functional and dynamical mean field theory},
  author={Wills, John M and Alouani, Mebarek and Andersson, Per and Delin, Anna and Eriksson, Olle and Grechnyev, Oleksiy},
  year={2010},
  publisher={Springer Science \& Business Media},
  url={https://link.springer.com/book/10.1007/978-3-642-15144-6}
}

@article{PhysRevB.61.8906,
  title = {First-principles calculations of magnetic interactions in correlated systems},
  author = {Katsnelson, M. I. and Lichtenstein, A. I.},
  journal = {Phys. Rev. B},
  volume = {61},
  issue = {13},
  pages = {8906--8912},
  numpages = {0},
  year = {2000},
  month = {Apr},
  publisher = {American Physical Society},
  doi = {10.1103/PhysRevB.61.8906},
  url = {https://link.aps.org/doi/10.1103/PhysRevB.61.8906}
}

@article{PhysRevB.68.104436,
  title = {First-principles relativistic study of spin waves in thin magnetic films},
  author = {Udvardi, L. and Szunyogh, L. and Palot\'as, K. and Weinberger, P.},
  journal = {Phys. Rev. B},
  volume = {68},
  issue = {10},
  pages = {104436},
  numpages = {11},
  year = {2003},
  month = {Sep},
  publisher = {American Physical Society},
  doi = {10.1103/PhysRevB.68.104436},
  url = {https://link.aps.org/doi/10.1103/PhysRevB.68.104436}
}

@article{PhysRevB.79.045209,
  title = {Anisotropic exchange coupling in diluted magnetic semiconductors: Ab initio spin-density functional theory},
  author = {Ebert, H. and Mankovsky, S.},
  journal = {Phys. Rev. B},
  volume = {79},
  issue = {4},
  pages = {045209},
  numpages = {6},
  year = {2009},
  month = {Jan},
  publisher = {American Physical Society},
  doi = {10.1103/PhysRevB.79.045209},
  url = {https://link.aps.org/doi/10.1103/PhysRevB.79.045209}
}

@article{fukui2005chern,
  title={Chern numbers in discretized Brillouin zone: Efficient method of computing (spin) Hall conductances},
  author={Fukui, Takahiro and Hatsugai, Yasuhiro and Suzuki, Hiroshi},
  journal={Journal of the Physical Society of Japan},
  volume={74},
  number={6},
  pages={1674--1677},
  year={2005},
  publisher={The Physical Society of Japan},
  url={https://doi.org/10.1143/JPSJ.74.1674}
}

@article{mukherjee2026skyrmion,
  title={Skyrmion manipulation and logic gate functionality in transition metal multilayers},
  author={Mukherjee, Tamali and Satya Narayana Murthy, V and Sadhukhan, Banasree},
  journal={Journal of Physics D: Applied Physics},
  volume={59},
  number={8},
  pages={085003},
  year={2026},
  publisher={IOP Publishing},
  url={https://iopscience.iop.org/article/10.1088/1361-6463/ae45b9/meta}
}

@article{yang2024fundamentals,
  title={Fundamentals and applications of the skyrmion Hall effect},
  author={Yang, Sheng and Zhao, Yuelei and Zhang, Xichao and Xing, Xiangjun and Du, Haifeng and Li, Xiaoguang and Mochizuki, Masahito and Xu, Xiaohong and {\AA}kerman, Johan and Zhou, Yan},
  journal={Applied Physics Reviews},
  volume={11},
  number={4},
  year={2024},
  publisher={AIP Publishing},
  url={https://doi.org/10.1063/5.0218280}
}

@article{sadhukhan2025engineering,
  title={Engineering skyrmion from spin spiral in transition metal multilayers},
  author={Sadhukhan, Banasree},
  journal={Journal of Physics: Condensed Matter},
  volume={37},
  number={9},
  pages={095801},
  year={2025},
  publisher={IOP Publishing},
  url={https://iopscience.iop.org/article/10.1088/1361-648X/ad9da8/meta}
}

@article{SciPostPhys.18.2.064,
	title = {Spin-lattice couplings and effect of displacements on magnetic interactions of a skyrmion system {PdFe/Ir(111)}},
	pages = {064},
	author = {Sadhukhan, Banasree and Bergman, Anders and Hellsvik, Johan and Thunström, Patrik and Delin, Anna},
	journal = {SciPost Phys.},
	volume = {18},
	year = {2025},
	publisher = {SciPost},
	doi = {10.21468/SciPostPhys.18.2.064},
	url = {https://scipost.org/10.21468/SciPostPhys.18.2.064}
}

@article{MUKHERJEE2025173036,
title = {Interplay between interfacial Dzyaloshinskii–Moriya interaction and magnetic anisotropy in 4d transition metal multilayers for skyrmion nucleation},
journal = {Journal of Magnetism and Magnetic Materials},
volume = {625},
pages = {173036},
year = {2025},
issn = {0304-8853},
doi = {https://doi.org/10.1016/j.jmmm.2025.173036},
url = {https://www.sciencedirect.com/science/article/pii/S0304885325002689},
author = {Tamali Mukherjee and Banasree Sadhukhan and V. Satya Narayana Murthy}
}

@article{PhysRevB.103.094410,
  title = {First-principles calculation of the Dzyaloshinskii-Moriya interaction: A Green's function approach},
  author = {Mahfouzi, Farzad and Kioussis, Nicholas},
  journal = {Phys. Rev. B},
  volume = {103},
  issue = {9},
  pages = {094410},
  numpages = {9},
  year = {2021},
  month = {Mar},
  publisher = {American Physical Society},
  doi = {10.1103/PhysRevB.103.094410},
  url = {https://link.aps.org/doi/10.1103/PhysRevB.103.094410}
}

@article{PhysRevB.105.104418,
  title = {Spin-lattice couplings in two-dimensional {CrI$_3$} from first-principles computations},
  author = {Sadhukhan, Banasree and Bergman, Anders and Kvashnin, Yaroslav O. and Hellsvik, Johan and Delin, Anna},
  journal = {Phys. Rev. B},
  volume = {105},
  issue = {10},
  pages = {104418},
  numpages = {10},
  year = {2022},
  month = {Mar},
  publisher = {American Physical Society},
  doi = {10.1103/PhysRevB.105.104418},
  url = {https://link.aps.org/doi/10.1103/PhysRevB.105.104418}
}

@article{PhysRevB.109.024420,
  title = {Mixed Bloch-N\'eel type skyrmions in a two-dimensional lattice},
  author = {He, Zhonglin and Dou, Kaiying and Du, Wenhui and Dai, Ying and Huang, Baibiao and Ma, Yandong},
  journal = {Phys. Rev. B},
  volume = {109},
  issue = {2},
  pages = {024420},
  numpages = {8},
  year = {2024},
  month = {Jan},
  publisher = {American Physical Society},
  doi = {10.1103/PhysRevB.109.024420},
  url = {https://link.aps.org/doi/10.1103/PhysRevB.109.024420}
}

@article{PhysRevB.107.L081110,
  title = {Effect of chirality imbalance on Hall transport of {PrRhC$_2$}},
  author = {Sadhukhan, Banasree and Nag, Tanay},
  journal = {Phys. Rev. B},
  volume = {107},
  issue = {8},
  pages = {L081110},
  numpages = {7},
  year = {2023},
  month = {Feb},
  publisher = {American Physical Society},
  doi = {10.1103/PhysRevB.107.L081110},
  url = {https://link.aps.org/doi/10.1103/PhysRevB.107.L081110}
}

@article{rkr4-p5n4,
  title = {Orbital-driven topological phase transition and planar Hall response in ternary telluride Weyl semimetals},
  author = {Sadhukhan, Banasree and Nag, Tanay},
  journal = {Phys. Rev. B},
  volume = {113},
  issue = {15},
  pages = {155130},
  numpages = {11},
  year = {2026},
  month = {Apr},
  publisher = {American Physical Society},
  doi = {10.1103/rkr4-p5n4},
  url = {https://link.aps.org/doi/10.1103/rkr4-p5n4}
}

@article{PhysRevB.110.174412,
  title = {Topological magnon in exchange frustration driven incommensurate spin spiral of kagome-lattice {YMn$_6$Sn$_6$}},
  author = {Sadhukhan, Banasree and Bergman, Anders and Thunstr\"om, Patrik and Lopez, Manuel Pereiro and Eriksson, Olle and Delin, Anna},
  journal = {Phys. Rev. B},
  volume = {110},
  issue = {17},
  pages = {174412},
  numpages = {7},
  year = {2024},
  month = {Nov},
  publisher = {American Physical Society},
  doi = {10.1103/PhysRevB.110.174412},
  url = {https://link.aps.org/doi/10.1103/PhysRevB.110.174412}
}

\section*{Appendix}

\subsection{Magnetic interactions}

\par Density functional theory (DFT) calculations for the electronic structure and magnetic interactions were carried out using the full-potential linear muffin-tin orbital (FP-LMTO) implementation in the RSPt code \cite{wills2010full}, employing both the local spin-density approximation (LSDA) and generalized gradient approximation (GGA) exchange-correlation functionals. A dense $18\times18\times18$ $k$-mesh was used for all DFT calculations. To accurately account for the electronic correlations associated with the Mn-$3d$ orbitals, we further combined spin-polarized LSDA with dynamical mean-field theory (DMFT) using the spin-polarized $T$-matrix fluctuation-exchange impurity solver as implemented in the RSPt code. For the LSDA+DMFT calculations, a reduced $12\times12\times12$ $k$-mesh was adopted to optimize the computational cost. In all cases, careful convergence tests with respect to the Brillouin-zone integration $k$-mesh were performed for both DFT and DMFT calculations.

\par The Heisenberg Hamiltonian describing the magnetic system is given by 
\bea
 \mathrm{H} = -\sum_{ij} \sum_{\{\alpha,\beta\}} e^{\alpha}_i \mathrm{J}^{\alpha \beta}_{ij} e^{\beta}_j 
\label{eq:HamS}
\eea
where $e^{\alpha}_i$ ($e^{\beta}_j$) is the $\alpha$ ($\beta$) component of the unitary vector pointing along the direction of the spin located at the site $i$ ($j$).  Considering   $\mathrm{J}^{\alpha \beta}_{ij}$ as a [3$\times$3] matrix, the isotropic (Heisenberg) part of the magnetic exchange interactions J$_{ij}$'s

and anti-symmetric Dzyaloshinskii–Moriya interactions (DMIs) D$_{ij}$'s are defined by 
\begin{eqnarray}
\mathrm{J}_{ij}= ({\mathrm{J}_{ij}}^{xx}+{\mathrm{J}_{ij}}^{yy}+{\mathrm{J}_{ij}}^{zz})/3,  \nonumber
\\ \nonumber
\mathrm{D}_{ij} = \mid \vec D_{ij} \mid = \sqrt{({\mathrm{D}_{ij}}^x)^2 + ({D_{ij}}^y)^2 + ({D_{ij}}^z)^2}, \\
\label{def-int}
\end{eqnarray}
where $\mid \vec D \mid $ is the magnitude of the DMI vector.  Here we used the convention of positive J$_{ij}$'s as ferromagnetic (FM) and negative J$_{ij}$'s as antiferromagnetic (AFM).

\par For a given material, the interaction parameters appearing in Eq.~\eqref{def-int} can be obtained using the magnetic force theorem within the framework of linear-response theory. This formalism was originally developed for isotropic Heisenberg exchange interactions in the absence of spin--orbit coupling (SOC) \cite{SECCHI201561, PhysRevB.61.8906}. The method relies on a second-order perturbative treatment of small deviations of the spin moments from the equilibrium magnetic configuration. Later, the approach was extended to include relativistic effects associated with SOC, thereby enabling the evaluation of the complete exchange interaction tensor $\mathrm{J}^{\alpha \beta}_{ij}$ \cite{PhysRevB.68.104436, PhysRevB.79.045209}.

\par Here we present a derivation of the formulae based on Green's functions formalism below.  We begin by perturbing the spin system by deviating the initial moments ($\vec e_0$) on a small angle $\vec{\delta \varphi}$ (the site index is omitted at the moment):
\bea
\vec e = \vec e_0 + \delta \vec e + \delta^2 \vec {e} = \vec e_0 + \bigl[ \vec{\delta \varphi} \times \vec e_0 \bigl] -\frac{1}{2} \vec e_0 (\vec {\delta \varphi})^2 \nonumber
\eea
Then one can write the Hamiltonian Eq.~\eqref{eq:HamS} of the perturbed system in terms of series in the order of $\vec \delta \varphi$:
\bea
\hat H' = \hat H^{(0)} + \hat H^{(1)} + \hat H^{(2)}. \nonumber
\eea
In the collinear limit,  all spins point along the same direction, which we set parallel to $Z$ axis. Then the tilting vectors have the following components:
\bea
\vec{\delta\varphi} = (\delta\varphi^x ; \delta\varphi^y ; 0)  \nonumber \\
\bigl[ \vec{\delta\varphi} \times \vec e_0 \bigl] = (\delta \varphi^y ; - \delta\varphi^x ; 0) \nonumber
\eea
Focusing on the energy contributions of the second order in $\vec{\delta \varphi}$ (i.e. $\hat H^{(2)}$), we obtain:
\bea
\mathrm{H}^{(2)} = -\sum_{i \ne j} \biggl( \mathrm{J}^{xx}_{ij} \delta\varphi_i^y \delta\varphi_j^y + \mathrm{J}^{yy}_{ij} \delta\varphi_i^x \delta\varphi_j^x - \mathrm{J}^{xy}_{ij} \delta\varphi_i^y \delta\varphi_j^x \nonumber \\ 
- \mathrm{J}^{yx}_{ij} \delta\varphi_i^x \delta\varphi_j^y -\frac12 \mathrm{J}^{zz}_{ij} ((\vec{\delta \varphi_i})^2 + (\vec{\delta \varphi_j})^2) \biggl) \nonumber
\eea
Then one can do the same perturbation for the electronic Hamiltonian ($\mathcal{H}$), which will become:
\bea
\hat{\mathcal{H}'} = \hat U^\dagger  \hat{\mathcal{H}} \hat U = \hat{\mathcal{H}}^{(0)} + \hat{\mathcal{H}}^{(1)} + \hat{\mathcal{H}}^{(2)}, \nonumber
\eea
where $\hat U=\exp{(i\vec{\delta\varphi} \hat{\vec{\sigma}}}/2)$ and $\hat{\vec{\sigma}}$ is the vector of Pauli matrices.
The corresponding terms proportional to $\vec{\delta\varphi}$ can be identified and mapped onto generalized Heisenberg model. The expressions for various components of $J^{\alpha\beta}_{ij}$ (Eq.~\eqref{eq:HamS})  are obtained as
\bea
\mathrm{J}^{xx}_{ij}= \frac{\text{T}}{4}\sum_p \text{Tr}_{L,m} \bigl[ \hat{\mathcal{H}}_i , \hat \sigma^{y} \bigl] G_{ij}(i\omega_p)   \bigl[ \hat{\mathcal{H}}_j , \hat \sigma^{y} \bigl] G_{ji} (i\omega_p) \nonumber \\
\mathrm{J}^{xy}_{ij}= -\frac{\text{T}}{4}\sum_p \text{Tr}_{L,m}  \bigl[ \hat{\mathcal{H}}_i , \hat \sigma^{y} \bigl] G_{ij}(i\omega_p)   \bigl[ \hat{\mathcal{H}}_j , \hat \sigma^{x} \bigl] G_{ji} (i\omega_p) \nonumber
\eea
and similar expressions are also for $\mathrm{J}{ij}^{yy}$ and $\mathrm{J}_{ij}^{yx}$.  The summation is done over the Matsubara frequencies ($\omega_p$) and the trace is over the orbital indices denoted by $m$.  The other components $\mathrm{J}^{xz}_{ij}$, $\mathrm{J}^{zx}_{ij}$, $\mathrm{J}^{yz}_{ij}$, $\mathrm{J}^{zy}_{ij}$ are not of the second order in the tilting angles.  Thus, for M $||$ $z$, only $D_z$ component (Eq.~\eqref{def-int}) can be computed, while $\mathrm{D}_x$ and $\mathrm{D}_y$ are extracted from two additional calculations with the magnetization pointing along $x$ and $y$, respectively. We have used the method described above to calculate the magnetic exchange interactions of YMn$_6$Sn$_6$ as implemented within RSPt code \cite{wills2010full}. 

\subsection{Monte Carlo annealing and spin dynamics simulations}
The magnetic ground state of the spin Hamiltonian was obtained using atomistic spin dynamics combined with Monte Carlo annealing, performed with the Uppsala Atomistic 
Spin Dynamics (UppASD) simulation package \cite{Eriksson2017}.   A two-dimensional lattice of size \(300 \times 300 \times 1\) with periodic boundary conditions applied in the in-plane directions was considered.

To obtain the relaxed initial spin configuration prior to the spin dynamics measurement,
a heat bath Monte Carlo (MC) annealing was performed.
Starting from a specified initial magnetic configuration and under a fixed external magnetic field, the system was annealed in 
six steps from $T = 100\,\mathrm{K}$ down to effectively 
zero kelvin ($T = 1 \times 10^{-5}\,\mathrm{K}$). 
The annealing schedule consisted of \(100{,}000\) MC steps at each of the temperatures \(100\,\mathrm{K}\), \(10\,\mathrm{K}\), \(1\,\mathrm{K}\), \(0.001\,\mathrm{K}\), and \(0.0001\,\mathrm{K}\), followed by a final equilibration of \(500{,}000\) MC steps at \(1\times10^{-5}\,\mathrm{K}\). In total, each annealing cycle involved \(1.0\times10^{6}\) MC steps.

Following the Monte Carlo annealing, the resulting spin configuration was further evolved using atomistic spin dynamics simulations in order to accurately determine the magnetic ground state. The spin dynamics simulations were performed by solving the Landau--Lifshitz--Gilbert (LLG) equation,

\begin{equation}
    \frac{d\mathbf{S}_i}{dt} = -\gamma_\mathrm{L} \mathbf{S}_i 
    \times \mathbf{B}_i - \gamma_\mathrm{L} \alpha 
    \mathbf{S}_i \times (\mathbf{S}_i \times \mathbf{B}_i)
\end{equation}

where \(\mathbf{S}_i\) denotes the local spin moment at site \(i\), $\mathbf{B}_i = -\partial \mathcal{H}_S / 
\partial(\mu_s \mathbf{S}_i)$ is the effective magnetic field acting on the spin, $\gamma_\mathrm{L} = \gamma/(1+\alpha^2)$ is the renormalized gyromagnetic ratio, and \(\alpha\) represents the damping parameter. In the present calculations, a damping constant of \(0.5\) and a time step of \(10^{-16}\,\mathrm{s}\) were used. Following the annealing stage, the system was evolved for an additional \(N_\mathrm{step}=1.0\times10^{6}\) spin-dynamics steps at a temperature of \(T=10^{-4}\,\mathrm{K}\), ensuring convergence to the magnetic ground state configuration.

\subsection{Skyrmion number}
Magnetic skyrmions possess a nontrivial topological spin configuration characterized by a finite winding of the spin texture in real space. The corresponding topological invariant is quantified by the skyrmion number, defined as \cite{Heinze2011-wt,Nagaosa2013}

\begin{equation}
N_{\mathrm{sk}}
=
\frac{1}{4\pi}
\int
\mathbf{S}\cdot
\left(
\frac{\partial \mathbf{S}}{\partial x}
\times
\frac{\partial \mathbf{S}}{\partial y}
\right)
\,dx\,dy.
\end{equation}

This quantity measures how many times the spin configuration wraps around the unit sphere. In numerical calculations on a discrete lattice, the continuous integral is replaced by a summation over lattice sites, while the spatial derivatives are computed using a central-difference five-point stencil scheme \cite{Ding2006}.
Fig.~\ref{sfig1} shows the evolution and convergence of the total energy and skyrmion number as a function of iteration steps.
\begin{figure}[ht]
\centering
\includegraphics[width=0.48\textwidth,angle=0]{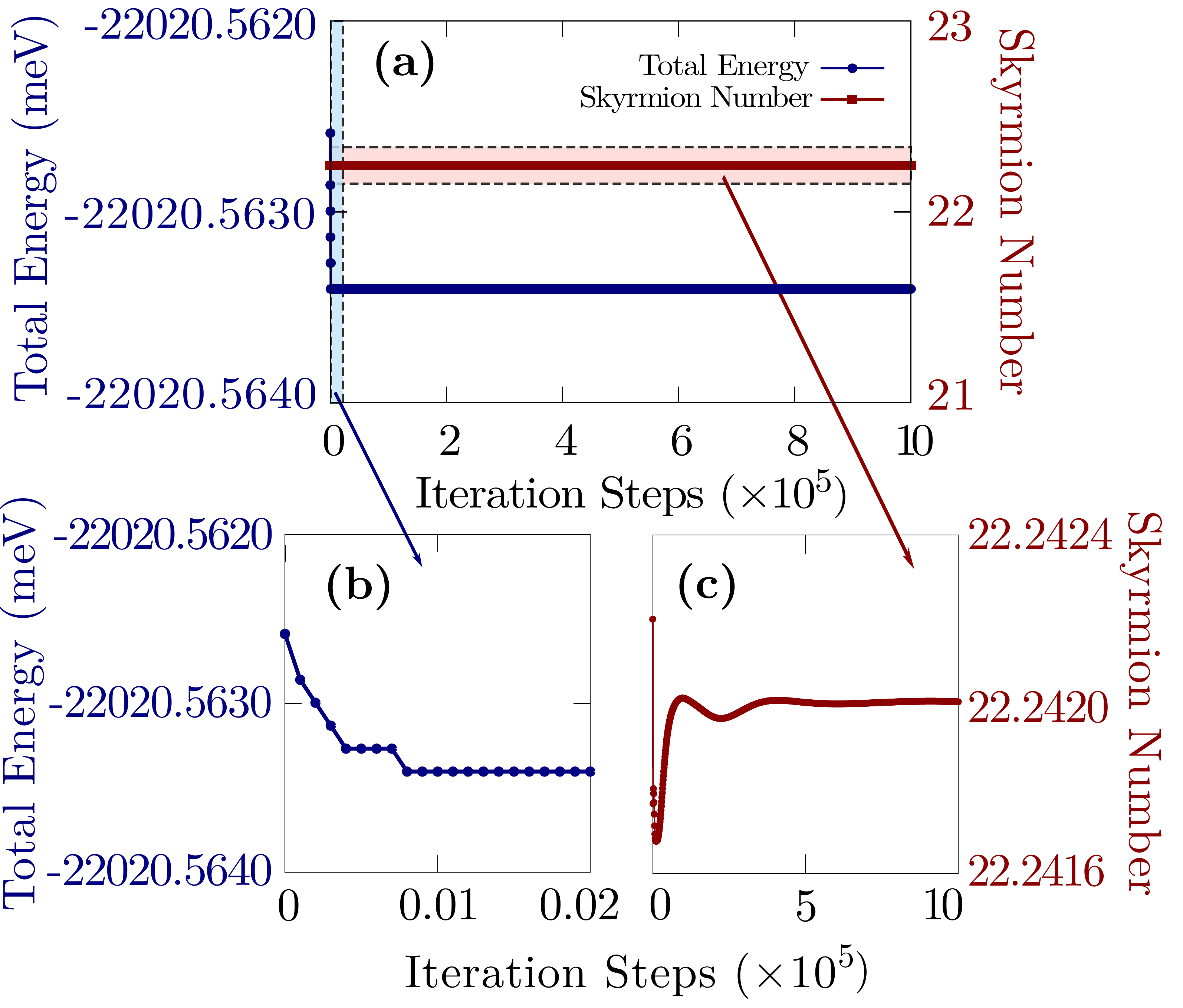}
\caption{(a) Total energy and skyrmion number as a function of iteration steps. The blue curve (left axis) and red curve (right axis) represent the total energy and skyrmion number, respectively. Enlarged views of the total-energy and skyrmion-number variations are shown in (b) and (c), respectively. }
\label{sfig1}
\end{figure}

\subsection{Spin correlation function}

To characterize the magnetic ordering and distinguish between different magnetic phases, we evaluate the spin correlation function in momentum space. This quantity is obtained from the Fourier transform of the real-space spin-spin correlations and is given by

\begin{equation}
S_{\mathbf{q}}
=
\frac{1}{N}
\sum_{ij}^{|\mathbf{r}_{ij}|<\delta}
\left\langle
\mathbf{S}_i \cdot \mathbf{S}_j
\right\rangle
e^{-i\mathbf{q}\cdot\mathbf{r}_{ij}},
\end{equation}

where \(\mathbf{S}_i\) represents the localized spin at lattice site \(i\), \(\mathbf{r}_{ij}\) denotes the distance between sites \(i\) and \(j\), and \(N\) is the total number of lattice sites. The cutoff radius \(\delta\) determines the spatial region over which correlations are included. Since \(S_{\mathbf{q}}\) corresponds directly to the magnetic Bragg intensity, it provides a convenient way to analyze magnetic ordering patterns and compare with neutron-scattering measurements.

\subsection{Topological Hall conductivity}

\begin{figure*}[ht]
\centering
\includegraphics[width=0.9\textwidth,angle=0]{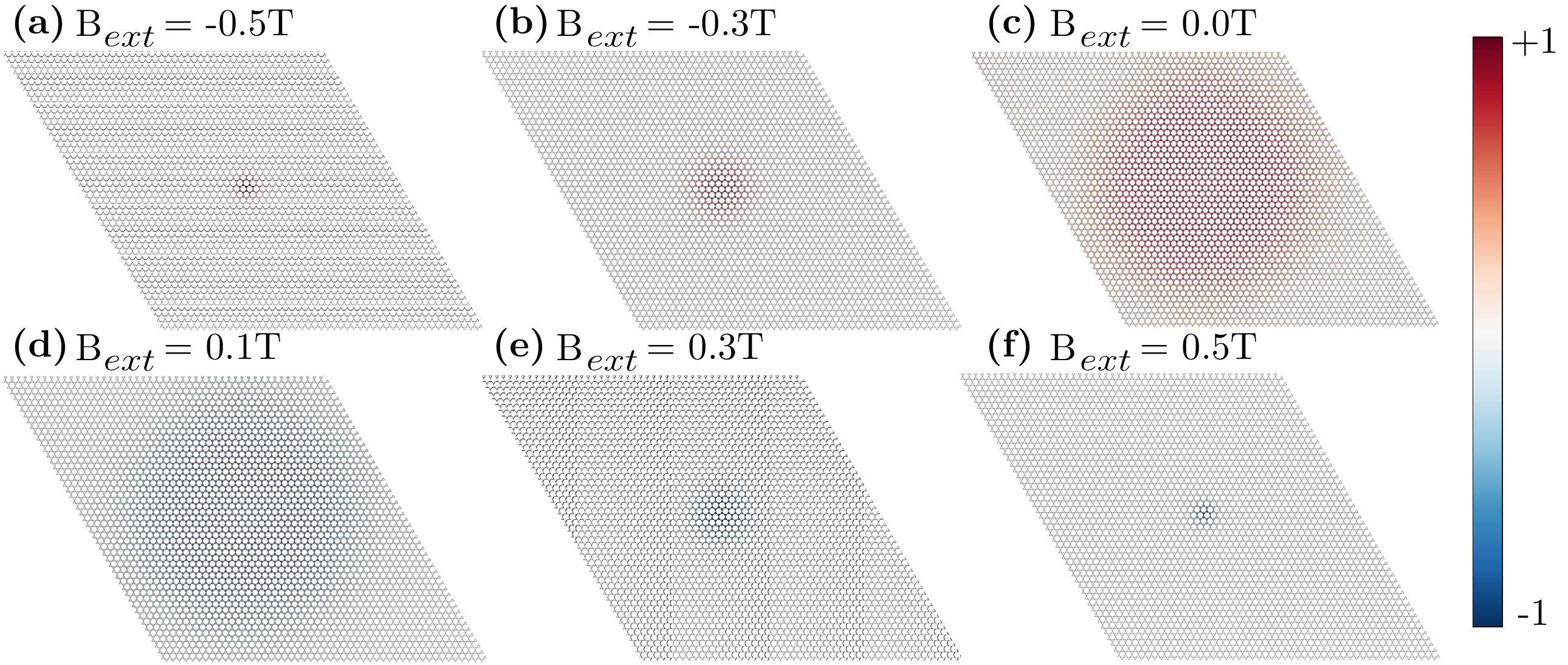}
\caption{Real-space resolved scalar spin chirality
    $\chi_{ijk} = \mathbf{S}_i \cdot (\mathbf{S}_j \times \mathbf{S}_k)$ 
    of the $50 \times 50 \times 1$ kagome lattice as a function of
external magnetic field $B_\mathrm{ext}$.}
\label{sfig2}
\end{figure*}

Fig.\ref{sfig2}(a)-(f) display the evolution of the local scalar spin 
chirality $\chi_{ijk}$ across the kagome lattice plaquettes 
as the external magnetic field $B_\mathrm{ext}$ is varied 
from $-0.5\,\mathrm{T}$ to $+0.5\,\mathrm{T}$. 
At $B_\mathrm{ext} = -0.5\,\mathrm{T}$, the chirality is 
confined to a compact, weakly positive core, indicative of 
an emerging skyrmion. As the magnitude of the negative field 
decreases (Fig.\ref{sfig2}(b)-(c)), the chirality core expands 
spatially and grows in amplitude, signaling the development 
of a fully formed skyrmion texture. At $B_\mathrm{ext} = 0.0\,\mathrm{T}$, 
the positive chirality fills the majority of the plaquettes. 
Upon crossing zero field, the chirality undergoes a sharp 
sign reversal (Fig.\ref{sfig2}(d)), producing a broad negative-chirality 
(blue) domain that subsequently shrinks and localizes as 
$B_\mathrm{ext}$ increases further (Fig.\ref{sfig2}(e)-(f)). 
This field-driven inversion of $\chi_{ijk}$ constitutes the 
real-space topological transition and is the microscopic origin 
of the sign change observed in the topological Hall conductivity.

\begin{figure*}[ht]
\centering
\includegraphics[width=1.0\textwidth,angle=0]{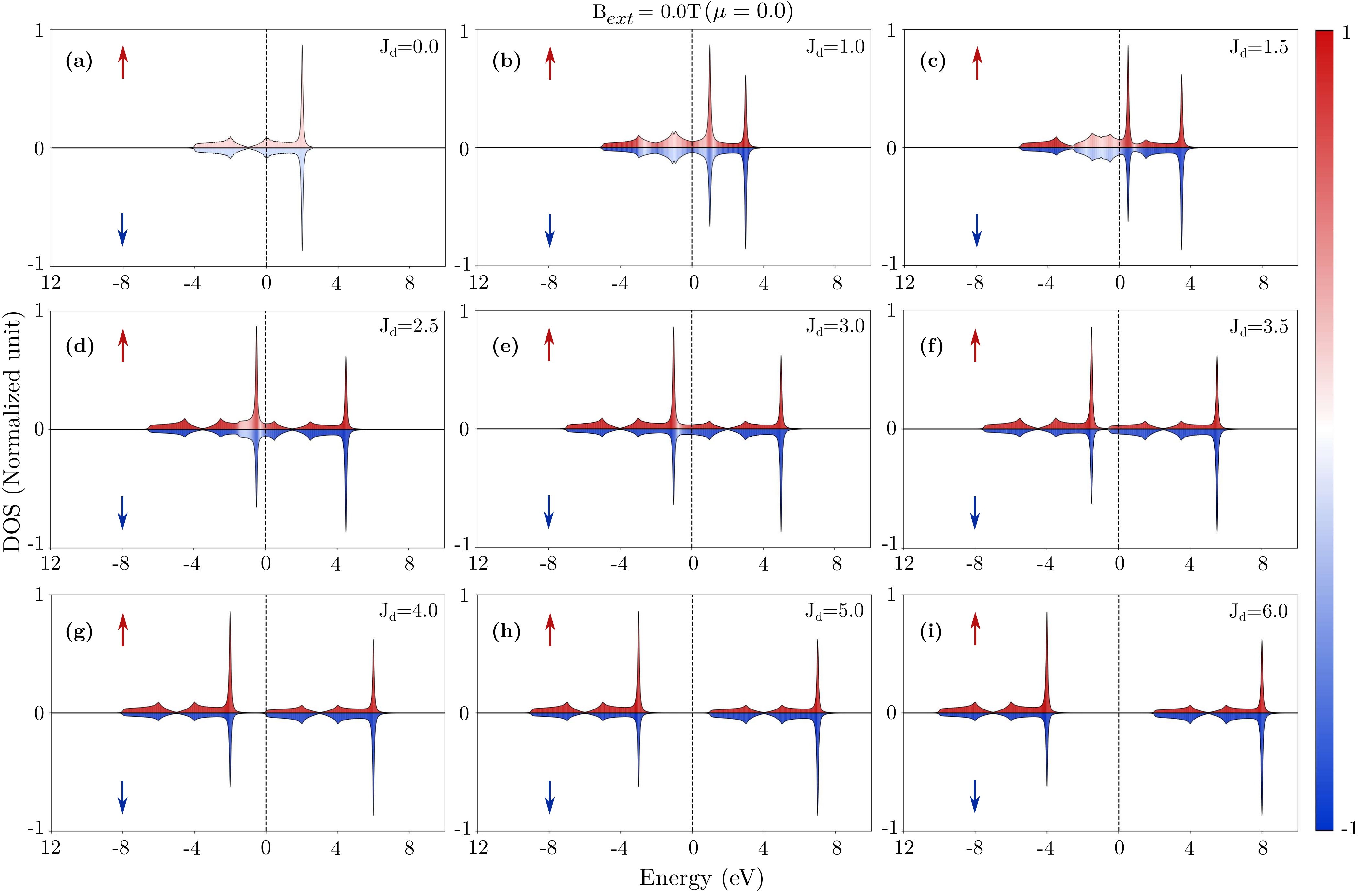}
\caption{Evolution of the spin-resolved density of states (DOS) with increasing $J_\mathrm{d}$ at zero external magnetic field ($B_\mathrm{ext}=0.0,\mathrm{T}$) and fixed chemical potential $\mu=0.0$. The DOS is plotted as a function of energy, with red (blue) shaded regions representing the spin-up (spin-down) contributions. The dashed vertical line denotes the chemical potential $\mu$.}
\label{sfig3}
\end{figure*}

Fig.\ref{sfig3}(a)-(i) shows the spin-resolved DOS at $B_\mathrm{ext}=0$  and $\mu = 0$ 
for $J_\mathrm{d}$ increasing from $0$ to $6.0$. Red (blue) lobes represent the spin-up (spin-down) density of states (DOS). 
At $J_\mathrm{d} = 0$ (Fig.\ref{sfig3}(a)), the spectrum 
reflects the bare spin-degenerate kagome bands. As $J_\mathrm{d}$ increases, 
the exchange field progressively lifts the spin degeneracy. The spin-up (spin-down) sub-bands shift
downward (upward) in energy, producing an increasingly pronounced spin-split spectrum. The flat-band singularity splits into two well-separated peaks, one in each spin channel. For $J_\mathrm{d} \geq 4.0$ (Fig.\ref{sfig3}(g)-(i)), the spin sub-bands are fully separated and the spectrum 
within each channel resembles the bare kagome dispersion. The progressive 
spin splitting with $J_\mathrm{d}$ governs where $\mu$ 
intersects the flat-band singularity and thus directly 
controls the magnitude and sign of $\sigma_{xy}$.

\begin{figure*}[ht]
\centering
\includegraphics[width=1.0\textwidth,angle=0]{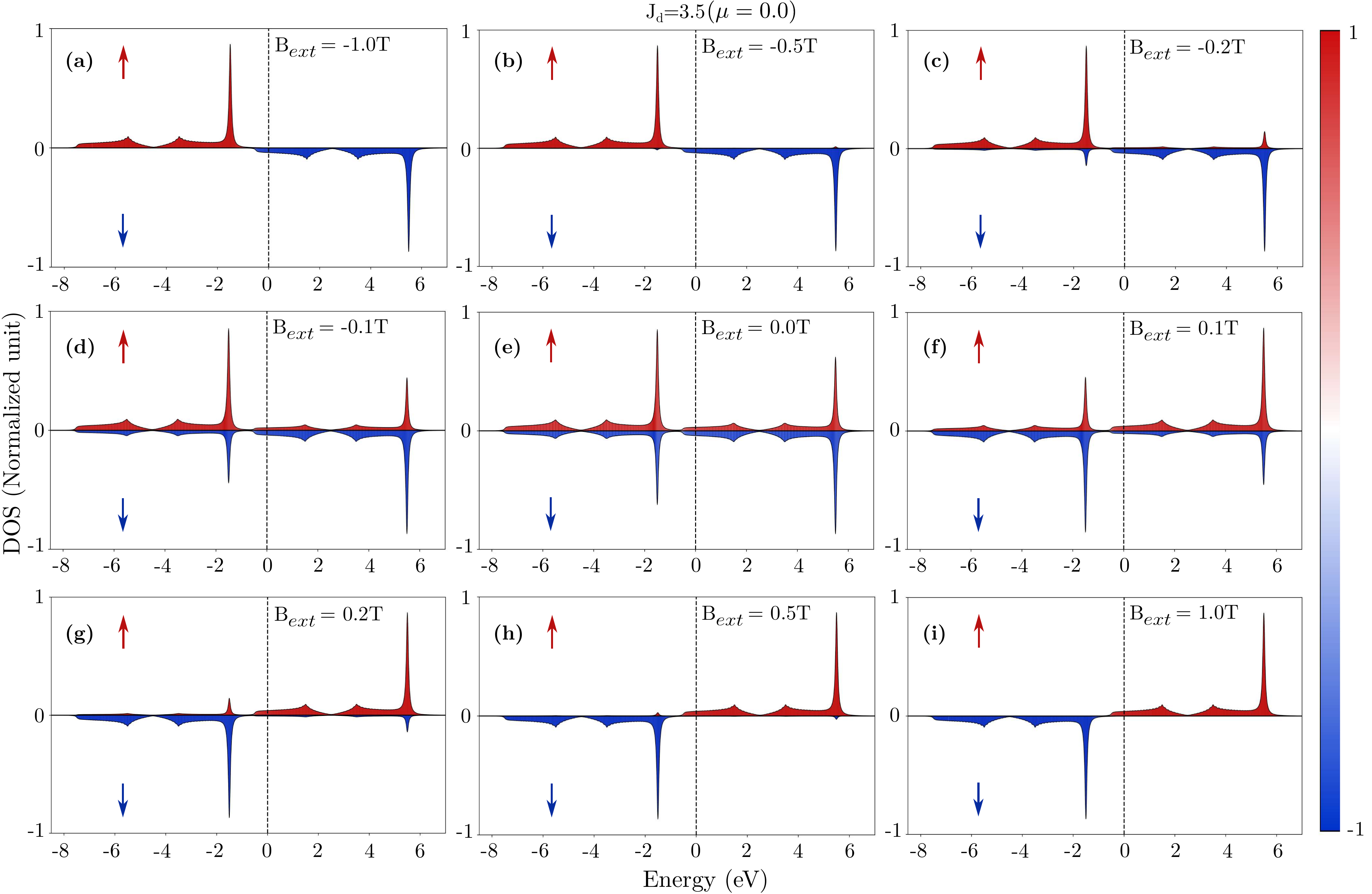}
\caption{Evolution of the spin-resolved density of states (DOS)  with increasing external magnetic fields $B_\mathrm{ext}$ for fixed $J_\mathrm{d}=3.5$ and chemical potential $\mu=0.0$. The DOS is plotted as a function of energy, with red (blue) shaded regions representing the spin-up (spin-down) contributions. The dashed vertical line denotes the chemical potential $\mu$.}
\label{sfig4}
\end{figure*}

Fig.\ref{sfig4}(a)-(i) shows the spin-resolved DOS at nine values of 
$B_\mathrm{ext}$ spanning the range $[-1.0, +1.0]\,\mathrm{T}$, 
at fixed $\mu = 0.0$ and $J_\mathrm{d} = 3.5$.  The red (positive $y$-axis) and blue (negative $y$-axis) lobes represent the spin-up ($\uparrow$) and spin-down ($\downarrow$) projected DOS respectively, normalized per site.
At large negative fields (Fig.\ref{sfig4}(a)-(c)), the spin-up 
sub-band dominates the occupied spectrum, 
while the spin-down sub-band is essentially unoccupied below $\mu$. 
Near zero field (Fig.\ref{sfig4}(d)-(f)), both spin channels contribute 
comparably to the DOS at $\mu$, reflecting the mixed spin 
character of the skyrmion state. At large positive fields 
(Fig.\ref{sfig4}(g)-(i)), the spectral weight is transferred entirely 
to the spin-down channel, signifying a complete reversal of the spin polarization. These field-induced changes in the spin-resolved electronic structure provide a clear spectroscopic signature of the underlying topological transition.

\begin{figure*}[ht]
\centering
\includegraphics[width=0.35\textwidth,angle=0]{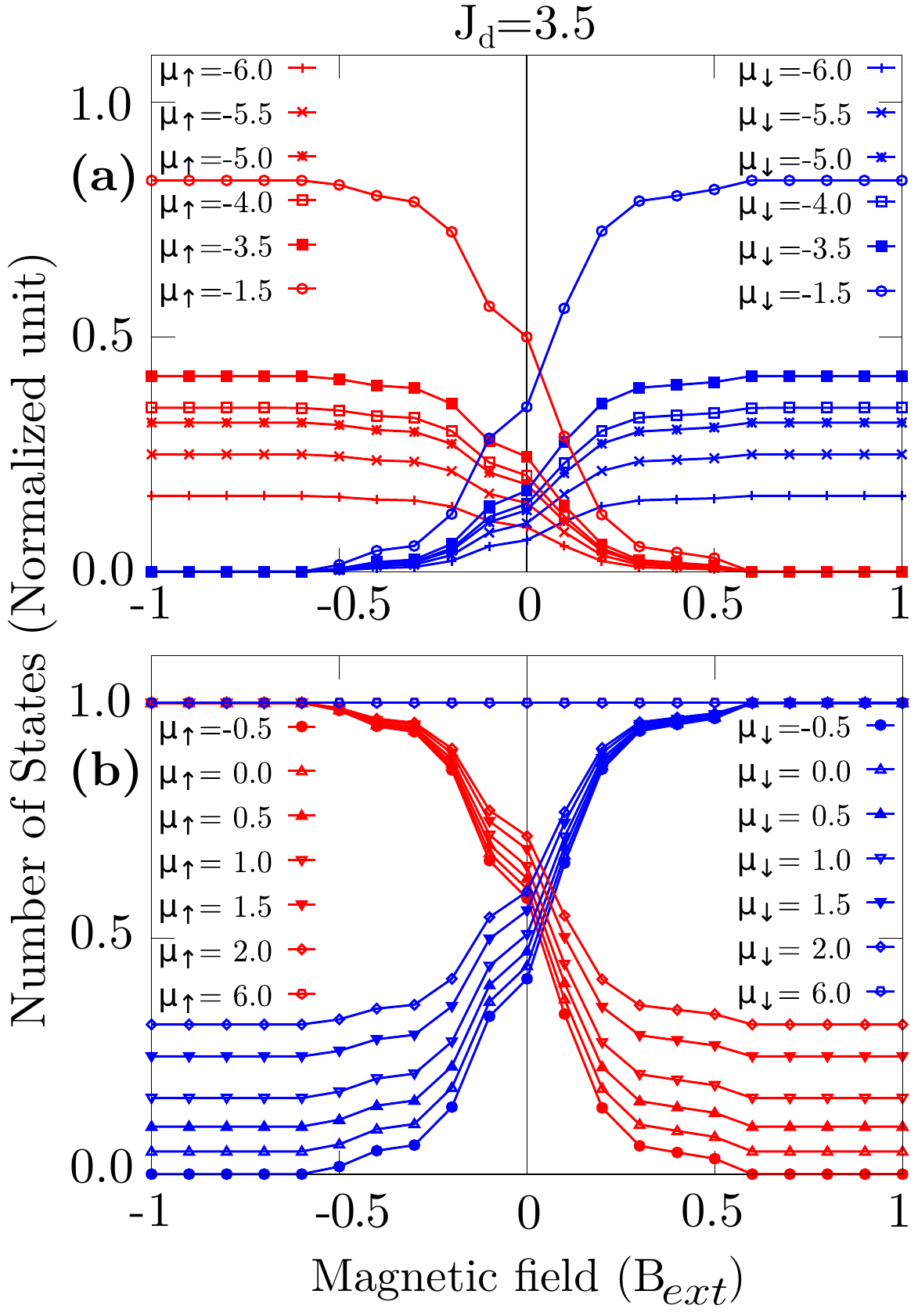}
\caption{Normalized spin-resolved spectral weight 
    (number of occupied states) as a function of $B_\mathrm{ext}$ 
    for $J_\mathrm{d} = 3.5$ and different chemical potentials. 
    Red (blue) curves correspond to spin-up (spin-down) channels. }
\label{sfig5}
\end{figure*}

In Fig.\ref{sfig5}(a)-(b) the normalized number of occupied spin-up (red) and 
spin-down (blue) states is shown as a function of 
$B_\mathrm{ext}$ for $J_\mathrm{d} = 3.5$ at multiple 
chemical potentials. In both panels the spin-up occupancy 
decreases monotonically with increasing $B_\mathrm{ext}$ 
while the spin-down occupancy increases, directly tracking 
the continuous rotation of the spin texture from a 
skyrmion configuration near $B_\mathrm{ext} \approx 0$ 
to a field-polarized ferromagnet at large $|B_\mathrm{ext}|$. Fig.\ref{sfig6} and \ref{sfig7} show Hall conductivity $\sigma_{xy}$ as a function of $B_\mathrm{ext}$ with chemical potential $\mu$ and Hunds's coupling J$_d$ respectively.

\par To resolve the microscopic, real-space origin of the geometric contribution to the Kubo formula, we
decompose the interband Berry curvature numerator onto individual nearest-neighbour bonds of the kagome
lattice, using the bond-resolved current operator along the $x$ and $y$ directions, respectively. The current operators are written as
\begin{equation}
\hat{J}_x=\sum_{(i,j)}\hat{J}_x^{\,ij}, \qquad
\hat{J}_y=\sum_{(i,j)}\hat{J}_y^{\,ij},
\end{equation}
where $(i,j)$ runs over nearest-neighbour bonds of the kagome lattice. Projecting one current operator
onto the bond basis while retaining the other in its global form defines the spatial distribution of the
Berry curvature numerator $\tilde{\Omega}_{mn}^{(ij)}$ using the bond-resolved current operator along the
$x$ and $y$ directions, shown in Fig.~\ref{sfig8} and Fig.~\ref{sfig9}, respectively,
\begin{align}
\tilde\Omega_{mn}^{(ij)} &= \mathrm{Im}\!\left[\langle m|\hat{J}_x^{\,ij}|n\rangle\langle n|\hat{J}_y|m\rangle\right],
\qquad \text{(bond-resolved along $x$)}, \\
\tilde{\Omega}_{mn}^{(ij)} &= \mathrm{Im}\!\left[\langle m|\hat{J}_x|n\rangle\langle n|\hat{J}_y^{\,ij}|m\rangle\right],
\qquad \text{(bond-resolved along $y$)},
\end{align}
evaluated for the interband transition between the highest occupied state $|m\rangle$ ($E_m\leq\mu$) and
the lowest unoccupied state $|n\rangle$ ($E_n>\mu$) at a given chemical potential $\mu$. Unlike the full
Kubo expression,  $\tilde{\Omega}_{mn}^{(ij)}$ does not carry the energy
denominator or the sum over all occupied--unoccupied pairs; each instead isolates how the geometric Hall
weight associated with a single, Fermi-level-adjacent transition is distributed in real space. For visualization, each bond-resolved weight is assigned to the corresponding bond midpoint and projected
along the physical bond direction. Denoting the unit bond vector by $\hat{e}_{ij}=\Delta\mathbf{r}_{ij}/
|\Delta\mathbf{r}_{ij}|$, the plotted quantity is $\tilde{\Omega}_{mn}^{(ij)}\,\hat{e}_{ij}$, so that each arrow lies along the kagome bond while its signed length and color encode the local magnitude
of the geometric Hall weight; the arrow orientation therefore reflects the bond geometry itself rather than
an independently defined current direction. $\tilde{\Omega}_{mn}^{(ij)}$ is
displayed on a symmetric diverging color scale centred at zero, with red (blue) denoting positive
(negative) weight.

The chemical potentials probed, $\mu_1=-6.0$, $\mu_2=-5.0$, $\mu_3=-4.0$, and $\mu_4=-3.0$ (marked in the
spin-resolved density of states in Fig.~\ref{sfig8}(a) and Fig.~\ref{sfig9}(a)), are chosen to straddle
the two principal singular features of the kagome band structure: the van Hove singularity, located near
$\mu\simeq-5.5$ and $-3.5$, and the Dirac point, located near $\mu\simeq-4.5$, with $\mu_1$ and $\mu_3$ sampling
the band filling on either side of these features. As $\mu$ is swept across the van Hove singularity and
subsequently across the Dirac point, the spatial distribution of the Berry curvature numerator
$\tilde{\Omega}_{mn}^{(ij)}$ using the bond-resolved current operator along both the $x$ and $y$ directions
(Figs.~\ref{sfig8} and \ref{sfig9}) undergoes a marked change in sign and spatial pattern, with the
dominant contribution switching between predominantly positive and predominantly negative weight. An analogous
sign change accompanies reversal of the external field, $B_{\mathrm{ext}}=\mp0.4$~T, at fixed $\mu$
(compare panels (b)--(e) with (f)--(i) of Fig.~\ref{sfig9}, and (b)--(e) with (f),(g) of Fig.~\ref{sfig8}).
Taken together, these results show that the interband phase coherence of the states closest to the Fermi level undergoes a significant reorganization as the chemical potential is tuned across the van Hove singularity and Dirac point. The resulting changes in the bond-resolved geometric response provide a real-space picture of how the electronic structure evolves with chemical potential and magnetic field, contributing to the observed dependence of the topological Hall response.

\begin{figure*}[ht]
\centering
\includegraphics[width=1.0\textwidth,angle=0]{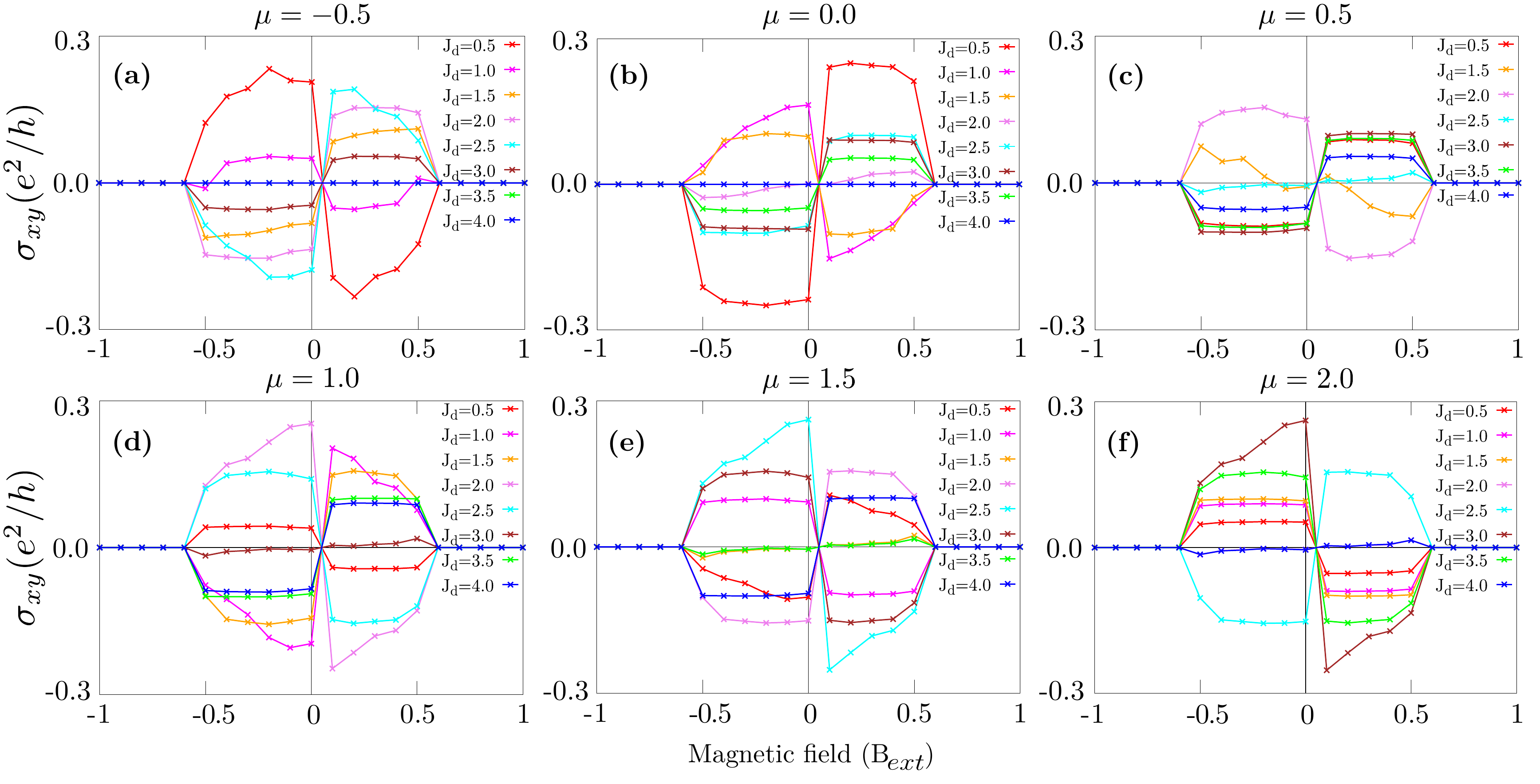}
\caption{Hall conductivity ($\sigma_{xy}$) ($e^2/h$) 
    as a function of $B_\mathrm{ext}$ for fixed chemical potential $\mu$ and varying $J_\mathrm{d}$ 
    from $0.5$ to $4.0$. Each panel in (a)-(f) corresponds to a distinct 
    value of $\mu$ as indicated.}
\label{sfig6}
\end{figure*}

\begin{figure*}[ht]
\centering
\includegraphics[width=1.05\textwidth,angle=0]{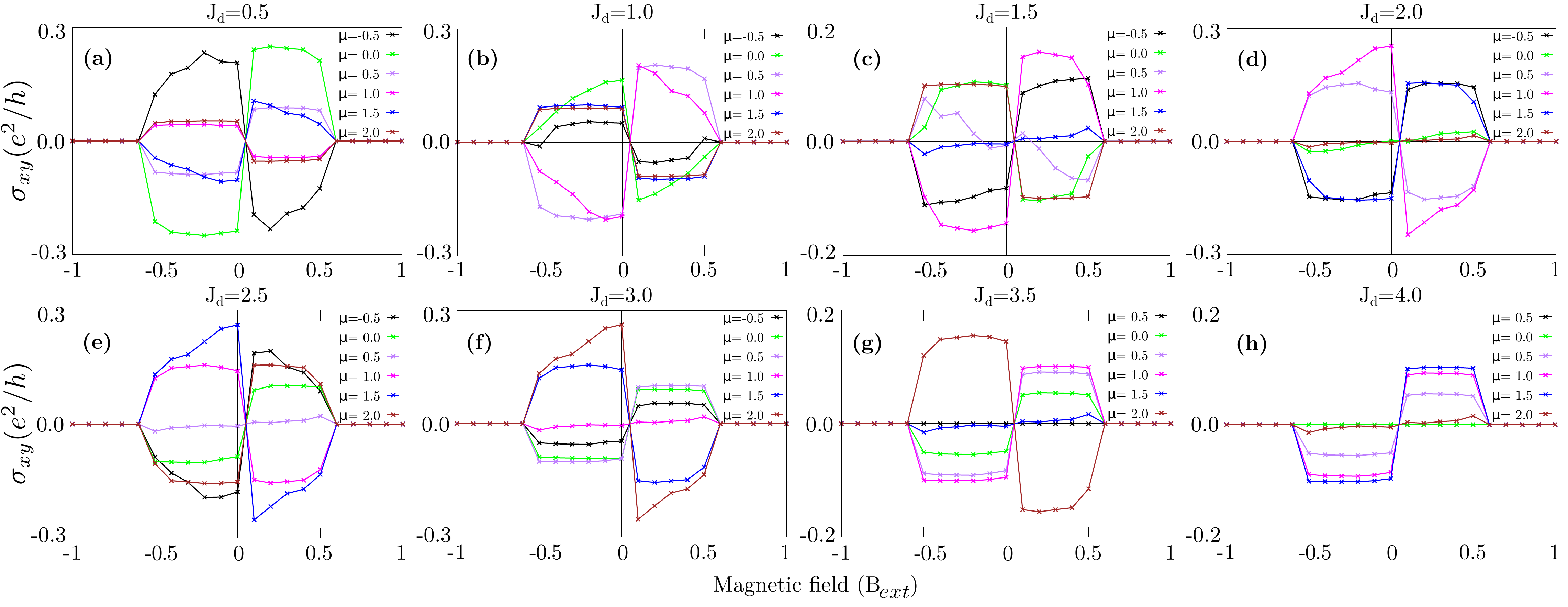}
\caption{Hall conductivity ($\sigma_{xy}$) ($e^2/h$) 
    as a function of $B_\mathrm{ext}$ for fixed  $J_\mathrm{d}$ and varying chemical potential 
    $\mu$ from $-0.5$ to $2.0$. Each panel in (a) -(h) corresponds to a 
    distinct value of $J_\mathrm{d}$ as indicated.}
\label{sfig7}
\end{figure*}

\begin{figure*}[ht]
\centering
\includegraphics[width=0.7\textwidth,angle=0]{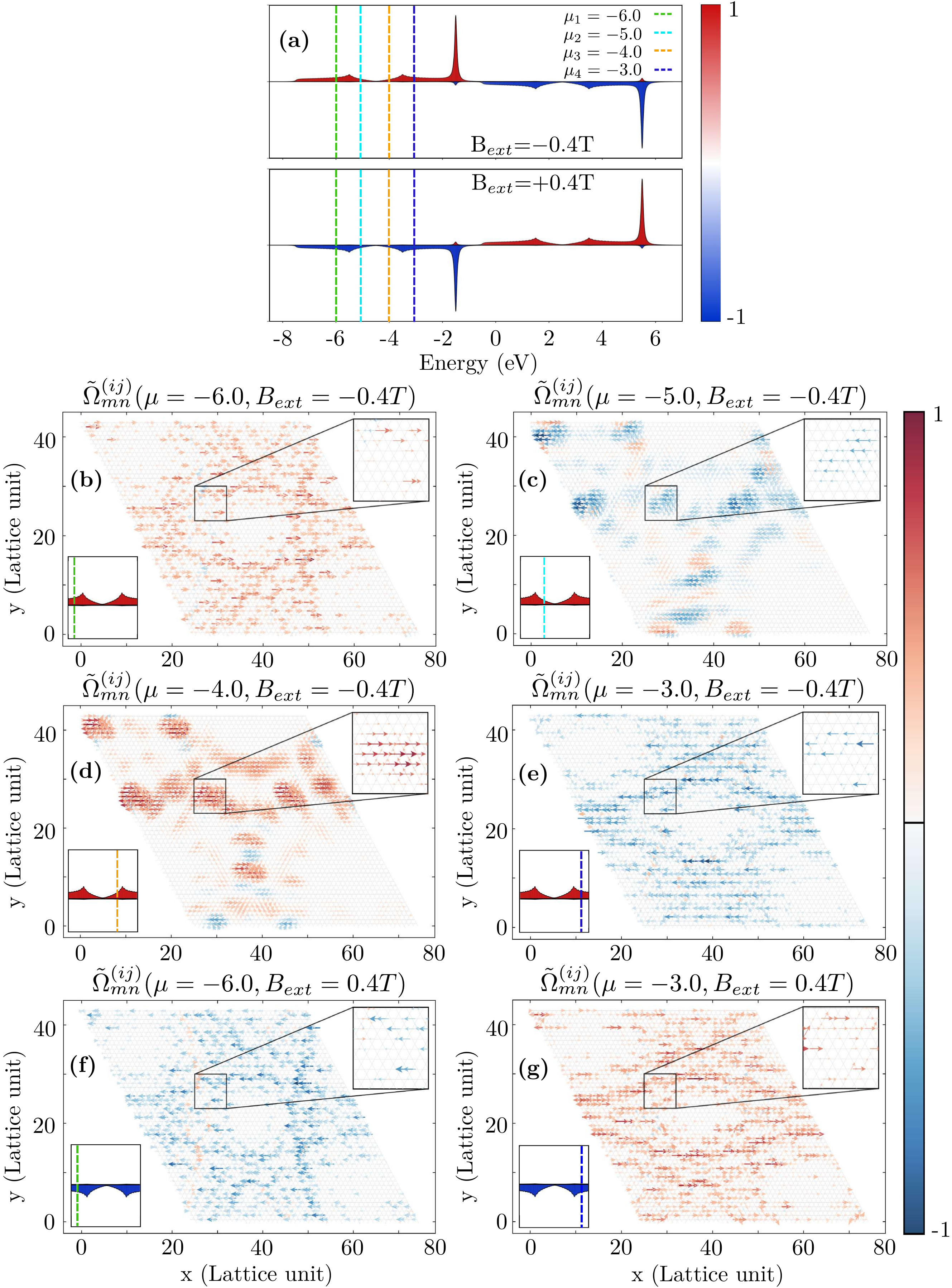}
\caption{Spatial distribution of Berry curvature numerator using the bond-resolved current operator along x direction $\tilde\Omega_{mn}^{(ij)}
= \mathrm{Im}
\left[
\langle m|\hat{J}_x^{\,ij}|n\rangle
\langle n|\hat{J}_y|m\rangle
\right]$  for the interband transition between the highest occupied and lowest unoccupied states at selected chemical potentials.  (a) Spin-resolved density of states for $B_{\mathrm{ext}}=-0.4$ T (top) and $B_{\mathrm{ext}}=+0.4$ T (bottom), with the chemical potentials $\mu=-6.0$, $-5.0$, $-4.0$, and $-3.0$ indicated by dashed vertical lines. (b)--(e) $\tilde\Omega_{mn}^{(ij)}$ at $B_{\mathrm{ext}}=-0.4$ T for $\mu=-6.0$, $-5.0$, $-4.0$, and $-3.0$, respectively. (f) and (g) corresponding distributions at $B_{\mathrm{ext}}=+0.4$ T for $\mu=-6.0$ and $-3.0$. Insets show magnified views of the selected regions together with the corresponding spin-resolved DOS highlighting the chosen chemical potential. Calculations are performed at $J_d=3.5$.}
\label{sfig8}
\end{figure*}

\begin{figure*}[ht]
\centering
\includegraphics[width=0.7\textwidth,angle=0]{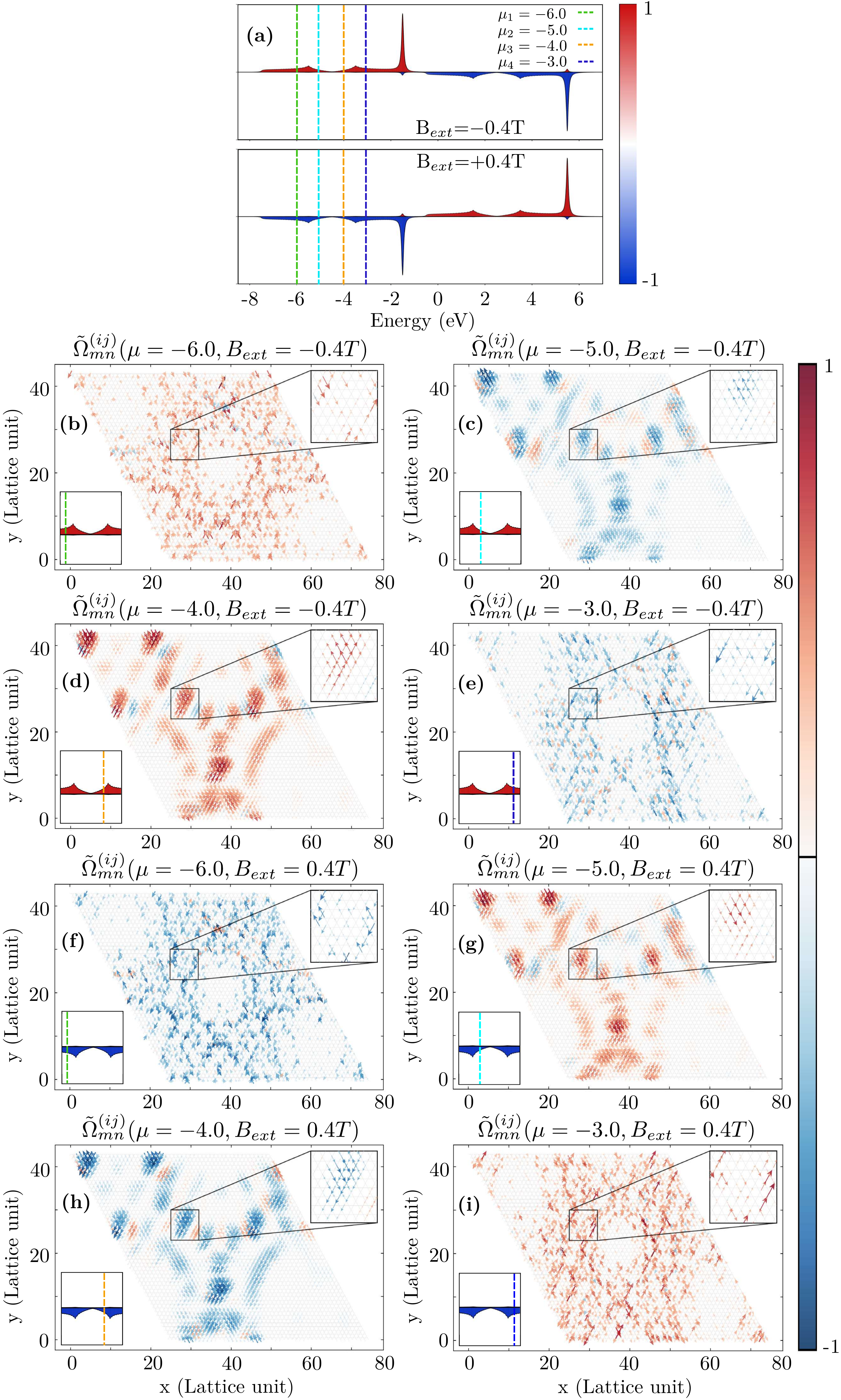}
\caption{Spatial distribution of Berry curvature numerator using the bond-resolved current operator along y direction $\tilde\Omega_{mn}^{(ij)}
= \mathrm{Im}
\left[
\langle m|\hat{J}_x|n\rangle
\langle n|\hat{J}_y^{\,ij}|m\rangle
\right]$  for the interband transition between the highest occupied and lowest unoccupied states at selected chemical potentials.  (a) Spin-resolved density of states for $B_{\mathrm{ext}}=-0.4$ T (top) and $B_{\mathrm{ext}}=+0.4$ T (bottom), with the chemical potentials $\mu=-6.0$, $-5.0$, $-4.0$, and $-3.0$ indicated by dashed vertical lines. (b)--(e) $\tilde\Omega_{mn}^{(ij)}$ at $B{\mathrm{ext}}=-0.4$ T for $\mu=-6.0$, $-5.0$, $-4.0$, and $-3.0$, respectively. (f)--(i) corresponding distributions at $B_{\mathrm{ext}}=+0.4$ T for same $\mu$ values . Insets show magnified views of the selected regions together with the corresponding spin-resolved DOS highlighting the chosen chemical potential. Calculations are performed at $J_d=3.5$.}
\label{sfig9}
\end{figure*}

\subsection{Momentum space Berry curvature and topological invariant}

The Chern numbers (topological invariants) and momenrum space Berry curvatures are calculated using a formalism based on Ref.~\cite{fukui2005chern} with the spin Hamiltonian described in main text. The Chern number of n'th band over a two-dimensional torus T$^2$ is given by
\begin{equation}
c_n=\frac{1}{2\pi i}\int_{T^2}d^2k\,F_{12}(k),
\label{CheNumCon}
\end{equation}
where the Berry field (Berry curvature)
strength~$F_{12}(k)$ and Berry connection~$A_\mu(k)$ ($\mu=1$, 2) are given by
\begin{eqnarray}
      &&F_{12}(k)=\partial_1A_2(k)-\partial_2A_1(k),
\nonumber\\
&&A_\mu(k)=\braOket{k}{\partial_\mu}{k},
\label{cont-f}
\end{eqnarray}
with $\ket{k}$ being a normalized wave function of the $n$th Bloch band
such that $H(k)\ket{k}=E_n(k)\ket{k}$.

Now we consider a two-dimensional Brillouin zone (BZ) where $\mu=1, 2$ could be $x,y,z$ consisting of discrete lattice points~$k_\ell$ given by 
($\ell=1$, \dots, $N_1N_2$) given by
\begin{equation}
   k_\ell=(k_{j_1},k_{j_2}),\quad
   k_{j_\mu}=\frac{2\pi j_\mu}{q_\mu N_\mu},\quad
   (j_\mu=0,\ldots,N_\mu-1).
\end{equation}
The Berry curvature is defined as 
\begin{eqnarray}
   &&\tilde F_{12}(k_\ell)\equiv
   \ln U_1(k_\ell)U_2(k_\ell+\hat1)U_1(k_\ell+\hat2)^{-1}U_2(k_\ell)^{-1},
\nonumber\\
   &&-\pi<\frac{1}{i}{\tilde F}_{12}(k_\ell)\leq\pi.
\label{FieStr}
\end{eqnarray}
where $\U(1)$ is the scalar product of the wave function of the $n$th
band at two consecutive reciprocal points in the BZ
\begin{equation}
U_\mu(k_\ell)\equiv \braket{k_\ell | k_\ell+\hatmu}/{\cal N}_\mu(k_\ell),
\label{seven}
\end{equation}
with ${\cal N}_\mu(k_\ell)=|\braket{k_\ell | k_\ell+\hatmu}|$.
Finally, a new Chern number for $n$th band is calculated by summing up the imaginary part of the Berry curvature for the discrete points on the BZ and it is defined as
\begin{equation}
   \tilde c_n\equiv\frac{1}{2\pi i}\sum_\ell\tilde F_{12}(k_\ell).
\end{equation}
We used 300$\times$300$\times$1  kmesh to calculate the Chern numbers and momentum space Berry curvature for YMn$_6$Sn$_6$  using the above formalism.

\end{document}